\documentclass[preprintnumbers,prd,floatfix,nofootinbib,superscriptaddress]{revtex4}

\usepackage[paperwidth=215.9mm,paperheight=279.4mm,centering,hmargin=2cm,vmargin=2.5cm]{geometry}

\usepackage[tbtags]{amsmath}  
\usepackage{tabularx}
\usepackage{amssymb}          
\usepackage{bm}               
\usepackage{graphicx}         
\usepackage[export]{adjustbox}
\usepackage{hhline,multirow}  
\usepackage{dcolumn}          
\usepackage{slashed}
\usepackage{datetime}
\usepackage{color}
\usepackage[dvipsnames]{xcolor}
\usepackage{slashed}
\usepackage{subfloat}
\usepackage{subeqnarray}

\usepackage{amscd}            
\usepackage{epsfig}
\usepackage{dcolumn}          
\usepackage[dvipsnames]{xcolor}
\usepackage[normalem]{ulem}
\usepackage{mathtools,cases}
\usepackage{comment}
\usepackage[utf8]{inputenc}

\RequirePackage{color}
\RequirePackage[colorlinks=true
,urlcolor=black
,anchorcolor=black
,citecolor=black
,filecolor=black
,linkcolor=black
,menucolor=black
,pagecolor=black
,linktocpage=black
,pdfproducer=medialab
,pdfa=true
]{hyperref}

\def\AAcom#1{{\bf  \textcolor{Red}{[AA: {#1}]}}}

\def\AScom#1{{\bf  \textcolor{Blue}{[AS: {#1}]}}}

\newcommand{\timeord}{{\cal T}}
\newcommand{\antitimeord}{{\cal \overline{T}}}
\newcommand{\colav}{\frac{\text{Tr}_c}{N_c}}

\newcommand{\Tr}{{\rm Tr}}
\newcommand{\vect}[1]{\boldsymbol{#1}}

\newcommand{\id}{{\mathbb{I}}}
\newcommand{\rom}{{|\Omega\rangle}}
\newcommand{\lom}{{\langle\Omega|}}
\newcommand{\llangle}{\langle\!\langle}
\newcommand{\rrangle}{\rangle\!\rangle}
\newcommand{\psibar}{{\overline{\psi}}}
\renewcommand{\i}{{\mathrm i}}

\newcommand{\nocontentsline}[3]{}
\newcommand{\tocless}[2]{\bgroup\let\addcontentsline=\nocontentsline#1{#2}\egroup}

\allowdisplaybreaks[2]

\begin{document}


\preprint{JLAB-THY-20-3193}

\title{On the connection between quark propagation and hadronization}

\author{Alberto Accardi}
\thanks{Electronic address: accardi@jlab.org - \href{https://orcid.org/0000-0002-2077-6557}{ORCID: 0000-0002-2077-6557}} 
\affiliation{Hampton University, Hampton, VA 23668, USA}
\affiliation{Theory Center, Thomas Jefferson National Accelerator Facility, 12000 Jefferson Avenue, Newport News, VA 23606, USA}

\author{Andrea Signori}
\thanks{Electronic address: asignori@jlab.org - \href{https://orcid.org/0000-0001-6640-9659}{ORCID: 0000-0001-6640-9659}} 
\affiliation{Dipartimento di Fisica, Universit\`a di Pavia, via Bassi 6, I-27100 Pavia, Italy}
\affiliation{INFN, Sezione di Pavia, via Bassi 6, I-27100 Pavia, Italy}
\affiliation{Theory Center, Thomas Jefferson National Accelerator Facility, 12000 Jefferson Avenue, Newport News, VA 23606, USA}

\begin{abstract}
We investigate the properties and structure of the recently discussed ``fully inclusive jet correlator'', namely, the gauge-invariant field correlator characterizing the final state hadrons produced by a free quark as this propagates in the vacuum. 
Working at the operator level, we connect this object to the single-hadron fragmentation correlator of a quark, and exploit a novel gauge invariant spectral decomposition technique to derive a complete set of momentum sum rules for quark fragmentation functions up to twist-3 level; known results are recovered, and new sum rules proposed. 
We then show how one can explicitly connect quark hadronization and dynamical quark mass generation by studying the inclusive jet's gauge-invariant mass term. This mass is, on the one hand, theoretically related to the integrated chiral-odd spectral function of the quark, and, on the other hand, is experimentally accessible through the $E$ and $\widetilde E$ twist-3 fragmentation function sum rules. 
Thus, measurements of these fragmentation functions in deep inelastic processes provide one with an experimental gateway into the dynamical generation of mass in Quantum Chromodynamics.
\end{abstract}

\date{\today}
\maketitle
\newpage
\tableofcontents

\newpage
\section{Introduction}
\label{s:intro}

One of the crucial properties of the strong force is confinement, namely the fact that color charged partons seemingly cannot exist as free particles outside of hadrons. As a consequence, any individual parton struck in a high-energy scattering process and extracted from its parent hadron must transform into at least one hadron -- in technical language, it must ``hadronize''.
During this process, a struck light quark, such as an up, down or strange, initially propagates as a high-energy but nearly massless colored particle, radiating by chromodynamic {\it bremsstrahlung} a number of other gluons and light quark-antiquark pairs (the radiation of heavy quarks such as the charm and the bottom is suppressed in proportion to their much higher mass, and can be ignored for the purposes of this discussion). 
Before reaching the experimental detectors, however, this system of colored, nearly massles particles will turn into a number of massive, color neutral hadrons such as pions, kaons and protons (with overall color charge conservation guaranteed, arguably, by soft final state interactions with the remnant of the parton's parent hadron). Hadronization is thus quite clearly and tightly connected to parton propagation, color charge neutralization, and dynamical generation of the mass, spin, and size of hadrons. However, the exact details of this parton-to-hadrons transition are poorly known. It is the purpose of this article to shed new light on these.

Unraveling hadronization dynamics is not only of fundamental importance to understand the emergence and nature of massive visible matter, but also an essential tool in hadron tomography studies at current and future facilities, including the 12 GeV program at Jefferson Lab~\cite{Dudek:2012vr} and a future US-based Electron-Ion Collider~\cite{Accardi:2012qut,Aidala:2020mzt}. For example, in Semi-Inclusive Deep Inelastic Scattering (SIDIS), measuring the transverse momentum of one of the final state hadrons can crucially provide a handle into the transverse motion of its parent quarks and gluons inside the hadron target~\cite{Angeles-Martinez:2015sea,Rogers:2015sqa,Bacchetta:2016ccz,Scimemi:2019mlf,Bacchetta:2019sam,Grewal:2020hoc,Bacchetta:2017gcc,Scimemi:2019cmh,Signori:2013mda,Anselmino:2013lza,Boglione:2014oea,Collins:2016hqq,Echevarria:2018qyi}. Understanding the hadronization mechanism is therefore critically important to quantitatively connect the initial, short-scale lepton-quark scattering hidden by confinement, with the measurable properties of hadrons as they hit the detectors.
%
%
%
%
Hadronization and, more in general, hadron structure are also very important for high-energy physics as they are among the biggest sources of uncertainty in the determination of Standard Model parameters~\cite{Webber:1999ui,Bozzi:2011ww,Quackenbush:2015yra,Bozzi:2015zja,CarloniCalame:2016ouw,Bacchetta:2018lna,Bozzi:2019vnl,Martinez:2019mwt} and the searches for physics beyond the Standard Model at the LHC~\cite{Gao:2017yyd,Rojo:2015acz}. 
Understanding hadronization is also essential for the study of cold and hot nuclear matter properties by means of jet quenching measurements in electron-nucleus and heavy ion collisions \cite{Accardi:2009qv, Arratia:2019vju}.

In high-energy collisions with a large four-momentum transfer, factorization theorems in Quantum Chromodynamics (QCD) allow one to separate the short-distance partonic scattering from the long-distance, non perturbative dynamics that binds the partons inside the target and detected particles~\cite{Parisi:1979se,Collins:1981uk,Collins:1989gx,Catani:2000vq,Becher:2010tm,GarciaEchevarria:2011rb,Collins:2011zzd,Echevarria:2012js, Chiu:2012ir,Rogers:2015sqa}. 
In this context, hadronization can be mapped -- and then utilized as a tool -- by means of fragmentation functions (FFs) that quantify the transmutation of a parton into one or more hadrons. FFs can be ``collinear'', namely, depending only on the ratio of the longitudinal momenta of the hadron and the parton, or ``trasverse-momentum-dependent'' (TMD), meaning they depend on both the longitudinal and transverse hadron momentum components.

The fragmentation functions can be determined by means of global QCD fits of hard semi-inclusive collisions. Collinear FFs for unpolarized hadrons are relatively well determined~\cite{Sato:2016wqj,Bertone:2017tyb,deFlorian:2014xna}, but there is currently no fit available for leading-twist polarized collinear FFs, such as the transversity FF $H_1$. The observation of a polarized hyperon in the final state could, however, shed light on the twist-3 collinear sector~\cite{Gamberg:2018fwy,Kanazawa:2015jxa}. In the transverse momentum sector, some information is available on the Collins TMD FF, among the polarized ones, as this involves polarized quarks but unpolarized hadrons~\cite{Kang:2015msa,Kang:2017btw,DAlesio:2017bvu,Anselmino:2013vqa,Anselmino:2015sxa,Anselmino:2015fty}. Lastly, while unpolarized TMD FFs are so far poorly known~\cite{Matevosyan:2011vj,Bentz:2016rav,Boglione:2017jlh}, present and forthcoming data from the {\tt BELLE} and {\tt BES-III} collaborations~\cite{Garzia:2016kqk,Seidl:2019jei,Seidl:2019FFtalk} will soon allow one to perform fits of these FFs, as well~\cite{Bacchetta:2015ora,Moffat:2019pci}. A comprehensive review on the theory and phenomenology of fragmentation functions, including di-hadron FFs and gluon FFs, can be found in Ref.~\cite{Metz:2016swz}.

The behavior of the fragmentation functions can be usefully constrained in a global QCD fit utilizing suitable sum rules~\cite{Jimenez-Delgado:2013sma}.
A number of sum rules for single-hadron FFs are  documented in literature~\cite{Collins:1981uw,Jaffe:1993xb,Mulders:1995dh,Schafer:1999kn,Bacchetta:2006tn,Meissner:2010cc,Accardi:2017pmi}, starting from the well known momentum sum rule for the unpolarized $D_1$ fragmentation function originally introduced in Ref.~\cite{Collins:1981uw}. A few have also been proposed for di-hadron FFs~\cite{Konishi:1979cb,deFlorian:2003cg,Majumder:2004br,Metz:2016swz}.
As we will see, however, the interest of FF sum rules also extends beyond their application to phenomenological fits, since a few of these are also sensitive to aspects of the non-perturbative QCD dynamics, such as the dynamics of mass generation.


The aim of this paper is to develop a field-theoretical formalism enabling us to take a fresh look at quark propagation and hadronization in the QCD vacuum.
Our strategy is to establish an operator-level master sum rule connecting the quark-to-hadron fragmentation correlator, that describes the transition of a quark into a hadron and an unobserved remnant ~\cite{Mulders:1995dh,Bacchetta:2006tn}, with the ``fully inclusive jet correlator'', that describes the fragmentation of a quark into an unobserved jet of particles~\cite{Sterman:1986aj,Collins:2007ph,Accardi:2008ne,Accardi:2017pmi}. 
We do this by generalizing the techniques utilized in Ref.~\cite{Meissner:2010cc}. 
We will then systematically exploit this correlator-level sum rule, and derive a complete set of sum rules for hadron spin independent fragmentation functions up to the twist-3 level. Results for selected Dirac structures have already been presented in Ref.~\cite{Accardi:2019luo}; in this work, that also provides full details of our approach, we complete the set of twist-3 sum rules (some of which generalize known results) and comment on their theoretical and phenomenological implications.

We would like to stress already here that, while the fully inclusive jet correlator also finds an application in the QCD factorization of, {\em e.g.}, inclusive DIS scattering at large values of the Bjorken $x$ variable~\cite{Sterman:1986aj,Becher:2006nr,Becher:2006mr,Collins:2007ph,Accardi:2008ne,Accardi:2017pmi,Accardi:2018gmh,Manohar:2003vb,Manohar:2005az,Chen:2006vd,Chay:2005rz,Chay:2013zya}, in this paper we consider this correlator as a theoretical object of intrinsic interest, and as a tool to derive the aforementioned sum rules, {\it independently} of any scattering process in which it may find application. 
In fact, as we will see, the inclusive jet correlator can be rewritten as the gauge invariant propagator of a color-averaged quark, and the generation of intermediate hadronic states analyzed in terms of the quark's K\"allen-Lehmann spectral functions. The dynamics of mass generation in the quark hadronization process can thus be explicitly connected in a gauge invariant way to the propagation of a quark in the QCD vacuum. 

In Section~\ref{s:jetcor} we perform a spectral analysis of the jet correlator, that will yield a gauge invariant decomposition in terms of the jet's momentum $k$, its mass $M_j$, and its virtuality $K_j^2$ plus terms associated with the Wilson line that renders the correlator gauge invariant. The starting point of this analysis is the convolutional spectral representation of the gauge invariant quark propagator proposed in Ref.~\cite{Accardi:2019luo}, see Eq.~\eqref{e:spectral_convolution}. 
We believe that this spectral representation can also find application beyond the present paper, for example 
in the study of the gauge independence of objects such as the virtuality-dependent parton distributions of Ref.~\cite{Radyushkin:2016hsy}, that are playing an increasingly important role in the direct lattice QCD calculation of PDFs in momentum space~\cite{Orginos:2017kos,Joo:2019jct}. 

All the coefficients in the inclusive jet correlator's decomposition are gauge invariant. In particular, this allows us to identify $M_j$ with, and propose a gauge invariant definition of, the mass of a dressed quark. This mass can be calculated in the light-cone gauge as an integral involving the chiral-odd spectral function of the quark propagator (see Section~\ref{ss:spectr_dec} and Section~\ref{ss:TMD_J_corr}), and can be considered as an order parameter for the dynamical breaking of chiral symmetry (see Section~\ref{ss:TMDjet_recap}).

In Section~\ref{s:1h_rules}, we derive the master sum rule connecting the unintegrated single-hadron fragmentation correlator to the inclusive jet correlator, see  Eq.~\eqref{e:master_sum_rule}, and from this obtain momentum sum rules for FFs up to twist 3 by suitable Dirac projections. These sum rules are summarized in Section~\ref{ss:sumrules_summary}, where we extensively comment on their theoretical and phenomenological implications. 
In particular, we find that the jet mass can be expressed as the sum of the current quark mass, $m$, and an interaction-dependent mass, $m^{\rm corr}$, which enter, respectively, at the right hand side of the sum rules for the collinear twist-3 $E$ and $\widetilde E$ FFs, see Section~\ref{ss:qgq_sumrules}. Measurements of these fragmentation functions, therefore, provide one with a concrete way to experimental probe the the mass generation mechanism in QCD, and to study the dynamical breaking of the chiral symmetry. Furthermore, the $E$ and $\widetilde E$ sum rules provide a way to separate the contribution of each hadron flavor to the overall jet mass, giving one even more insight on these processes.
%

Finally, in Section~\ref{s:conclusions} we summarize the results and discuss possible extensions of our work, and in the appendices 
we provide details about our conventions and the Lorentz transformations of the fragmentation and inclusive jet correlators.



\section{The fully inclusive jet correlator and its spectral decomposition}
\label{s:jetcor}

\subsection{The inclusive jet correlator}

Let us start by considering the unintegrated 
inclusive quark-to-jet correlator~\cite{Sterman:1986aj,Chen:2006vd,Collins:2007ph,Accardi:2008ne,Accardi:2017pmi,Accardi:2019luo} 
\begin{align}
  \label{e:invariant_quark_correlator}
  \Xi_{ij}(k;w) = \text{Disc} \int \frac{d^4 \xi}{(2\pi)^4} e^{\i k \cdot \xi} \,
    \colav\, \lom 
    \big[\, \timeord\, W_1(\infty,\xi;w) \psi_i(\xi) \big]\, 
    \big[\, \antitimeord\, \psibar_j(0) W_2(0,\infty;w) \big] 
    \rom \ ,  
\end{align}
where $\rom$ is the interacting vacuum state of QCD, $\psi$ the quark field, $W_{1,2}$ are Wilson lines that ensure the gauge invariance of the correlator, and $w$ is an external vector that determines the direction of their paths, as discussed in detail later. $\timeord$ represents the time ordering operator for the fields whereas $\antitimeord$ represents the anti time ordering operator~\cite{Collins:2011zzd,Echevarria:2016scs}, and for sake of brevity we omit the flavor index of the quark fields and of $\Xi$ (but all the results in this paper should be understood to be flavor-dependent). The color trace of the correlator will be shortly discussed in detail.

\begin{figure}[tb] 
\centering
\begin{tabular}{ccc}
\includegraphics[width=0.32\linewidth,valign=b]{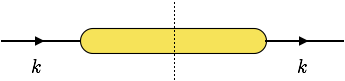}
& \hspace{2cm} &	
\includegraphics[width=0.32\linewidth,valign=b]{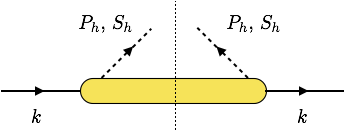}
\\ & & \\ 
(a) & & (b)
\end{tabular}
\caption{
Diagrammatic interpretations of (a) the fully inclusive jet correlator of Eqs.~\eqref{e:invariant_quark_correlator} and \eqref{e:invariant_quark_correlator_W}, and (b) the single-hadron fragmentation correlator \eqref{e:1hDelta_corr}. The black solid line corresponds to the hadronizing quark with momentum $k$, the black dashed line to the produced hadron with momentum $P_h$ and spin $S_h$. The yellow blob corresponds to the unobserved hadronization products. The vertical thin dashed line is the cut that puts the (unobserved) particles on the mass shell (see Appendix~\ref{a:conv}).
}
\label{f:cut_diagrams}
\end{figure}

The definition~\eqref{e:invariant_quark_correlator} clarifies and refines the definitions previously advanced in Refs.~\cite{Accardi:2008ne,Accardi:2017pmi,Accardi:2019luo}. 
In particular, with the present definition, the jet correlator naturally emerges at the right hand side of the master sum rule~\eqref{e:master_sum_rule}, that we will derive in the next section and connects the single inclusive quark fragmentation process to the propagation  and inclusive fragmentation of a quark discussed in this Section. 
A diagrammatic interpretation of Eq.~\eqref{e:invariant_quark_correlator} is given in Figure~\ref{f:cut_diagrams}(a), where the vertical cut line represents the discontinuity, ``Disc'', in Eq.~\eqref{e:invariant_quark_correlator}, or, in other words, a sum over all quark hadronization products~\cite{Sterman:1986aj}. In fact, inserting a completeness between the square brackets in Eq.~\eqref{e:invariant_quark_correlator}, one can interpret the jet correlator as the square of the sum of all possible quark-to-hadron transition amplitudes. 

The color-averaging of the initial-state quark, implemented as Tr$_c[\dots]/N_c$, has a crucial role, since it mimicks the color neutralization that has to take place in order to be able to consider the discontinuity, which corresponds to having on-shell intermediate states (see, {\it e.g.}, Ref.~\cite{Roberts:2015lja}, Section~2.4.). Furthermore, the color average is essential for the spectral representation of the jet correlator to be developed in Section~\ref{ss:spectr_dec}.  Finally, color averaging is also important in view of the sum rule discussion of Section~\ref{s:1h_rules} and, more in general, for using the inclusive jet correlator in factorization theorems at large Bjorken $x$~\cite{Chen:2006vd}. When calculating Dirac traces, we will also average over the quark's polarization states as detailed in Appendix~\ref{a:conv}. 

On the physics side, the correlator $\Xi$
captures the hadronization of a quark including {\em all} the products of the hadronization process. We call this the ``fully inclusive'' jet correlator\footnote{Other names used in the literature for the same correlator are ``jet distribution"~\cite{Sterman:1986aj} and ``jet factor"~\cite{Collins:2007ph}. Its Dirac projections have also been called  ``(final state) jet functions"~\cite{Accardi:2008ne,Chen:2006vd,Chay:2013zya}.} 
in order to stress that none of the jet's constituents is actually reconstructed -- hence the absence of a definition for a jet axis and radius, contrary to other semi-inclusive definition of jets. In the following, when using for simplicity the term ``jet''(or ``inclusive jet'') correlator, we will always refer to this fully inclusive jet correlator.
%
The inclusiveness of $\Xi$ will also be evident when relating this to the correlator for the hadronization of a quark into a single hadron by the sum rule we will prove in Section~\ref{s:1h_rules}. Finally, when inserting $\Xi$ in a DIS diagram~\cite{Accardi:2008ne,Accardi:2017pmi}, which can be justified at large values of Bjorken $x$ where 4-momentum conservation limits the amount of transverse momentum available to final state hadrons \cite{Sterman:1986aj,Chen:2006vd,Accardi:2018gmh}, the jet in question can be identified with the current jet. 

It is also interesting to remark that, taking into account the properties of the color trace and after a specific choice for the path of the Wilson line (to be discussed next), the correlator $\Xi$ can be expressed as the discontinuity of the gauge-invariant quark propagator, whose spectral decomposition has been studied in Ref.~\cite{Yamagishi:1986bj} for the case of a straight Wilson line connecting $0$ to $\xi$. In this paper, we will discuss instead the spectral representation for the case of Wilson lines running along staple-like contours -- which are the natural paths arising in QCD factorization theorems -- and use this in applications involving correlators integrated along one light-cone direction.

\subsubsection{Wilson line structure}
\label{sss:link_structure}

We work in a reference frame specified by two light-like unit vectors $n_+$ and $n_-$ such that $n_+^2 = n_-^2 = 0$ and $n_+\cdot n_- = 1$. Any other vector $a^\mu$ can then be specified in light-cone coordinates as $a=[a^-,a^+,\vect{a}_T]$, with $a^\pm = a\cdot n_\mp$, and $\vect{a}_T$ the 2-dimensional coordinates of $a$ transverse with respect to the $(n_+,n_-)$ plane, see Appendix~\ref{a:conv}. We also assume the quark to be highly boosted in the negative $z$ direction, so that the minus component of its momentum is dominant, $k^- \gg |\vect{k}_T| \gg k^+$. 
When the jet correlator is included in the calculation of a physical process, $n_+$ and $n_-$ can be determined by the kinematics of that process; for example, in DIS one can choose these to be coplanar with the hadron and virtual photon momenta, with $n_-$ aligned with the latter \cite{Accardi:2008ne,Accardi:2017pmi}. However, in this paper we study the jet correlator as a theoretical object in its own right. 

\begin{figure}
\centering
\begin{tabular}{ccccc}
\includegraphics[width=0.31\textwidth]{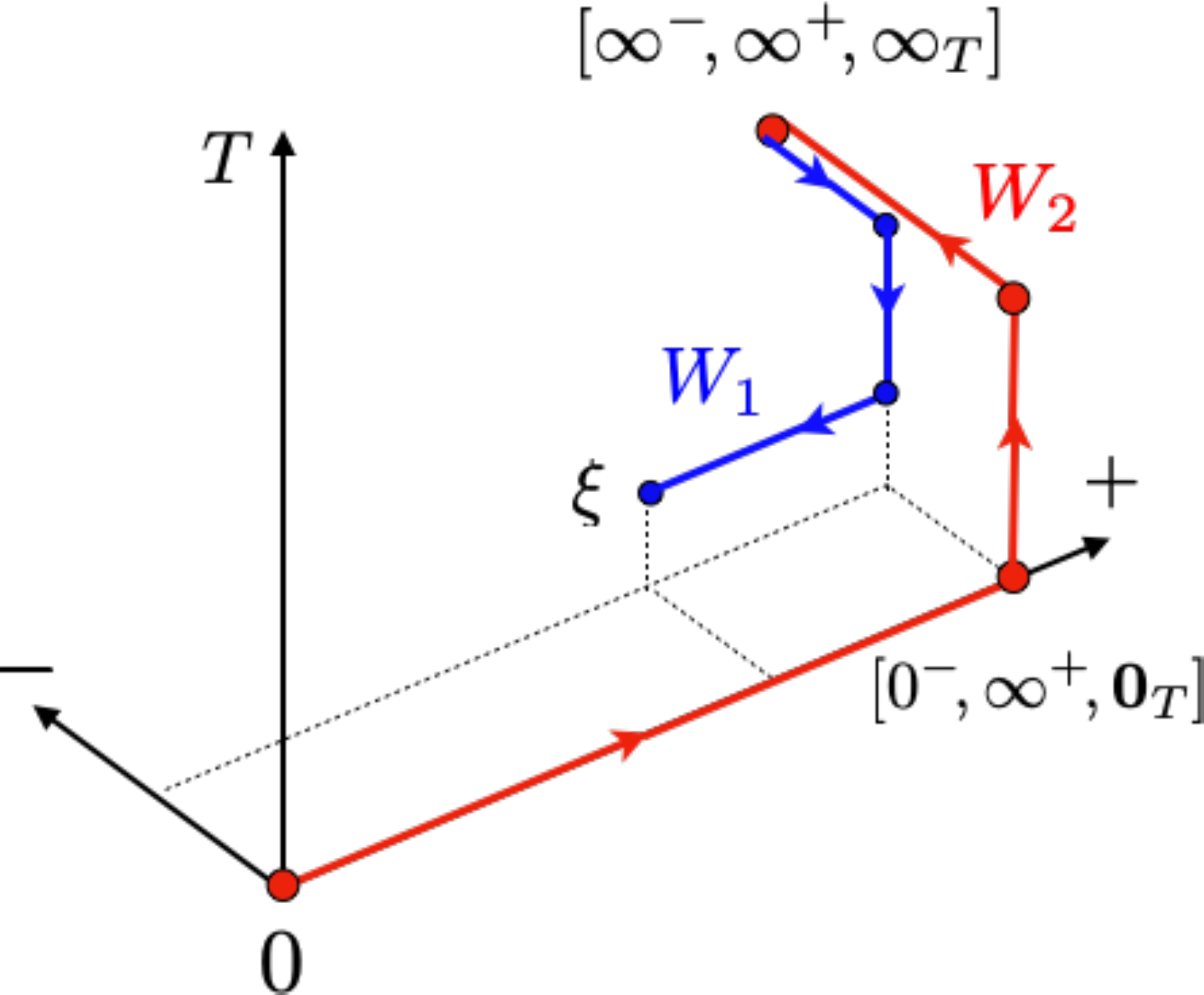}
& \hspace{0.15cm} &
\includegraphics[width=0.31\textwidth]{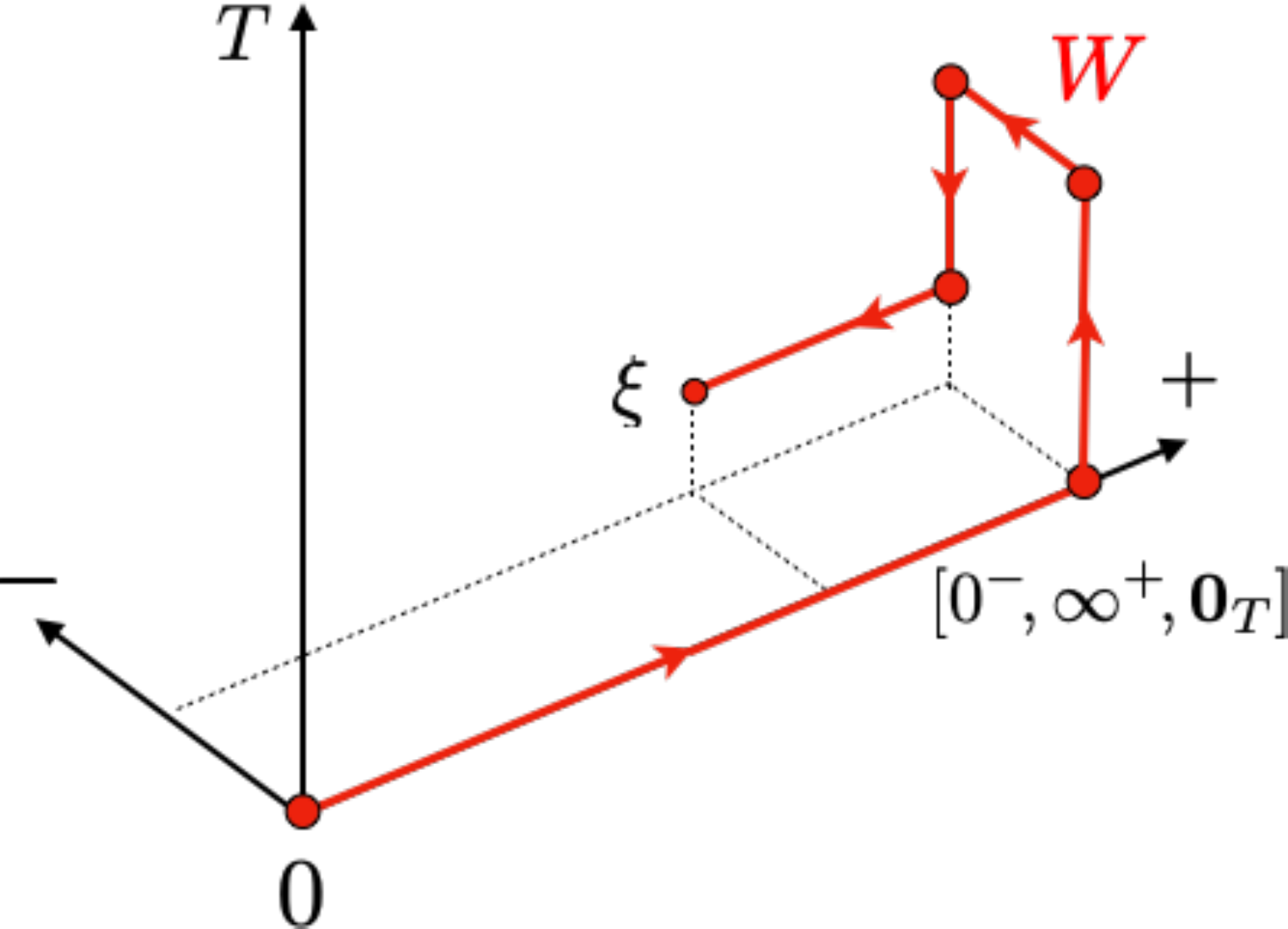}
& \hspace{0.15cm} &
\includegraphics[width=0.31\textwidth]{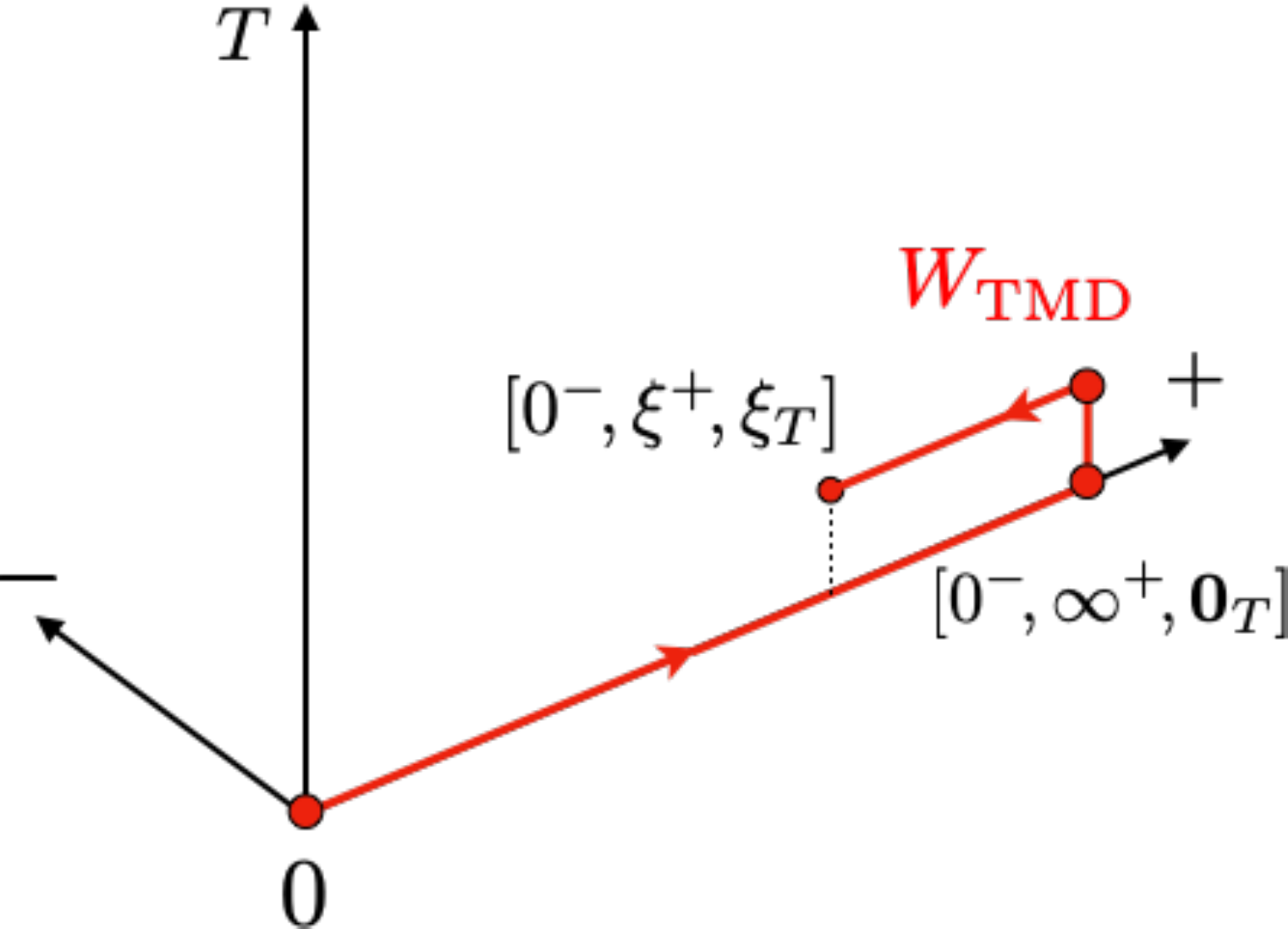}
\\
\quad\ (a) & & \quad\ (b) & & \quad\ (c)  
\\ 
\end{tabular}
\caption{Staple-like Wilson line paths for the jet correlator $\Xi(k;w=n_+)$: 
(a) $W_2(0,\infty;w=n_+)$ (red) and $W_1(\infty,\xi;w=n_+)$ (blue); 
(b) the combined Wilson line $W(0,\xi;w=n_+)$; 
(c) the TMD Wilson line $W_{TMD}(\xi^+,\vect{\xi}_T)$.
}
\label{f:W_paths}
\end{figure}

We restrict our attention to a class of Wilson lines that reproduces the paths determined by the TMD factorization theorems for quark-initiated hard scattering processes, when $\Xi$ is integrated over the sub-dominant momentum components $k^+$~\cite{Collins:2008ht,Collins:2011zzd} (or even in the fully unintegrated factorization proposed in Ref.~\cite{Collins:2007ph}). 
For simplicity, we also restrict the discussion to the case $w=n_+$, even though no substantial impediment arises in the treatment of slightly off-the-light-cone Wilson lines. 
To be specific, 
we take $W_2$ to run first from 0 to infinity along the plus light-cone direction,   
then to infinity in the transverse plane, 
and eventually to infinity again along the minus light-cone direction, see Figure~\ref{f:W_paths}(a).
Analogously, $W_1$  
runs from infinity backwards in the minus direction until it reaches $[\xi^-,\infty^+,\vect{\infty}_T$], 
then in the transverse direction until $[\xi^-,\infty^+,\vect{\xi}_T]$, 
and eventually reaching $\xi = [\xi^-,\xi^+,\vect{\xi}_T]$ along the plus direction 
(see Figure~\ref{f:W_paths}(a)).  
Explicitly, 
\begin{align}
\label{e:W2_antitime}
W_2(0,\infty;n_+) & =\, 
{\cal U}_{n_+}[0^-, 0^+, \vect{0}_T; 0^-, \infty^+, \vect{0}_T]\,\,
{\cal U}_{v_T}[0^-, \infty^+, \vect{0}_T; 0^-, \infty^+, \vect{\infty}_T]\, \, 
{\cal U}_{n_-}[0^-, \infty^+, \vect{\infty}_T; \infty^-, \infty^+, \vect{\infty}_T]\, ,
\\
\label{e:W1_time}
W_1(\infty,\xi;n_+) &=\,
{\cal U}_{n_-}[\infty^-, \infty^+, \vect{\infty}_T; \xi^-, \infty^+, \vect{\infty}_T]\,\, 
{\cal U}_{v_T}[\xi^-, \infty^+, \vect{\infty}_T; \xi^-, \infty^+, \vect{\xi}_T]\,\,
{\cal U}_{n_+}[\xi^-, \infty^+, \vect{\xi}_T; \xi^-, \xi^+, \vect{\xi}_T]\, , 
\end{align}
%
%
with  ${\cal U}_v$ representing the Wilson operator along a half-line starting from $a$ in the $v$ direction, {\it i.e.},
\begin{align}
  {\cal U}_v[a;\infty] = {\cal P} \exp \Big( -\i g \int_0^\infty ds\,  v^\mu A_\mu(a+sv) \Big) \, ,
\end{align}
where ${\cal P}$ denotes the path-ordering operator, and the square brackets emphasizes the straightness of the path. We will comment upon other possible choices of paths below.

From the definitions~\eqref{e:W2_antitime} and \eqref{e:W1_time}, one can see that $W_2$ is automatically anti time ordered, and $W_1$, time ordered. Namely, with our choice of paths, $\antitimeord {\cal P} \equiv {\cal P}$ and $\timeord {\cal P} \equiv {\cal P}$. 
Owing to this specific choice for the Wilson lines and thanks to the color trace and to the absence of intermediate states, we can also perform a cyclic permutation of the fields in Eq.~\eqref{e:invariant_quark_correlator} and combine the two Wilson operators $W_{1,2}$ into the single, staple-like operator 
\begin{align}
  W(0,\xi;n_+) = W_2(0,\infty;n_+) W_1(\infty,\xi;n_+) 
\end{align}
illustrated in Figure~\ref{f:W_paths}(b), so that
\begin{align}
  \label{e:invariant_quark_correlator_W}
  \Xi_{ij}(k;n_+) = \text{Disc} \int \frac{d^4 \xi}{(2\pi)^4} e^{\i k \cdot \xi} \,
  \colav\, \lom \psi_i(\xi) \psibar_j(0) W(0,\xi;n_+) \rom \, .
\end{align}
The jet correlator can thus also be written as the discontinuity of a gauge invariant quark propagator. 

Upon setting $\xi^-=0$, that corresponds to integrating the correlator over the $k^+$ component, we obtain the staple-like $W_{TMD}$ Wilson line  
utilized to define TMD distributions: 
\begin{align}
\label{e:W_np_TMD}
  W_{TMD}(\xi^+, \vect{\xi}_T) & \equiv W(0,\xi;n_+)|_{\xi^-=0} \nonumber \\
  & = 
    {\cal U}_{n_+}[0^-, 0^+, \vect{0}_T; 0^-, \infty^+, \vect{0}_T] \ 
    {\cal U}_{v_T}[0^-, \infty^+, \vect{0}_T; 0^-, \infty^+, \vect{\xi}_T] \
    {\cal U}_{n_+} [0^-, \infty^+, \vect{\xi}_T; 0^-, \xi^+, \vect{\xi}_T] \ , 
\end{align}
see Figure~\ref{f:W_paths}(c).  
Note that, even without choosing a specific form for $W_{1,2}$, the integration over $k^+$ renders the (anti-)time ordering on the light cone  equivalent to the path ordering, since $\xi^+$ becomes proportional to the time variable. 
Setting also $\vect{\xi}_T=0$, which corresponds to furthermore integrating $\Xi$ over the transverse momentum $\vect{k}_T$, one would finally obtain the collinear Wilson line  $W_{coll}(\xi^+) \equiv W_{TMD}(\xi^+,\vect{0}_T)={\cal U}[0^-, 0^+, \vect{0}_T; 0^-, \xi^+, \vect{0}_T]$, running along the light-like segment from $0^+$ to $\xi^+$.

It is important to remark that the results discussed in this paper hold true also for a larger class of Wilson lines. In order to drop the $\timeord$ and $\antitimeord$ time-ordering operators and rewrite the jet correlator as a gauge invariant quark propagator, one only needs to ensure that the time variable increases along the path that goes from 0 to $\infty$, and decreases when going from $\infty$ to $\xi$. This clearly restricts the class of Wilson lines available, but not overmuch. 
For example, one could 
consider off-the-light-cone lines, with $w \neq n_+$ slightly tilted away from the plus light-cone direction, 
or adopt L-shaped lines reaching infinity along $n_+$ and then moving to infinity simultaneously along the minus and transverse directions. 
Furthermore, we will need $W$ in Eq.~\eqref{e:invariant_quark_correlator_W} to reduce to the identity matrix when $\xi \to 0$, hence in that limit the path of the Wilson line should not contain loops. The straight line and the staple-like Wilson line that enter in collinear and TMD factorization belong to this category. 


\subsubsection{Dirac structure}
\label{sss:dirac_structure}

The correlator $\Xi$ can be decomposed on a basis of Dirac matrices,
$
  \big\{ \id, \gamma_5, \gamma^\mu, \gamma^\mu \gamma_5,
    \i\sigma^{\mu\nu} \gamma_5   \big\} ,
$
where $\i$ is the imaginary unit, $\id$ the identity matrix, 
$\gamma_5 = \i \epsilon^{\mu\nu\rho\sigma}\gamma_\mu\gamma_\nu\gamma_\rho\gamma_\sigma$,  $\epsilon^{\mu\nu\rho\sigma}$ the totally antisymmetric tensor of rank 4, 
and $\sigma^{\mu\nu} = (\i/2) [ \gamma^\mu,\gamma^\nu ]$. 
Assuming invariance under Lorentz and parity transformations, we can parametrize $\Xi$ as:
\begin{align}
  \Xi(k;n_+) = \Lambda A_1 \id + A_3 \slashed{k}
    + \frac{\Lambda^2}{k \cdot n_+} B_1 \slashed{n}_+
    + \frac{\Lambda}{k \cdot n_+} B_3 \sigma_{\mu\nu} k^\mu n_+^\nu \ .
  \label{e:jet_ampl}
\end{align}
The amplitudes
$A_i$ and $B_i$ are, in principle, functions of all the Lorentz scalars that one can build with the available Lorentz vectors, $k \cdot n_+$ and $k^2$~\cite{Mulders:1995dh,Bacchetta:2000jk,Boer:2016xqr}. 
The vector $n_+$ is available as the vector specifying the direction of the Wilson line.
The subscripts differ from the conventions discussed in Refs.~\cite{Accardi:2017pmi,Accardi:2018gmh}, but have been chosen such that they match the customary decomposition of the fragmentation correlator to be discussed in Section~\ref{ss:1h_inclusive_FF}. 
We have also introduced a power-counting scale $\Lambda = O(\Lambda_{\text{QCD}})$
that defines an ``operational'' twist expansion for the correlator in powers of $\Lambda/k^-$, where we assume a large boost in the $k^-$ direction so that $k^+ = (k^2+\vect{k}_T^2)/2k^- \ll |\vect{k}_T| \ll k^-$, with $\vect{k}_T \sim O(\Lambda)$ and $k^+,k^2\sim O(\Lambda^2)$. 
This expansion is analogous to that used in Ref.~\cite{Bacchetta:2004zf,Bacchetta:2006tn} for the single-hadron fragmentation correlator, and further discussed in Section~\ref{ss:1h_inclusive_FF}. Its relation with the rigorous twist expansion of local operators in the Operator Product Expansion formalism is discussed, {\it e.g..}, in Ref.~\cite{Jaffe:1996zw}. 

The twist expansion of the inclusive jet correlator can be made explicit by writing Eq.~\eqref{e:jet_ampl} in terms of the light-cone Dirac matrices $\gamma^\pm = \slashed{n}_\mp$:
\begin{align}
  \Xi(k;n_+) = k^-  A_3 \, \gamma^+
    + \Lambda\left( A_1\, \id
      + A_3\, \frac{\slashed{k}_T}{\Lambda}
      + \frac{\i}{2} B_3\, [\gamma^+,\gamma^-] \right)
    + \frac{\Lambda^2}{k^-} \left( B_1\, \gamma^-   
      + A_3 \frac{k^2+\vect{k}_T^2}{2\Lambda^2}\, \gamma^-  
      + B_3 \frac{\i}{2\Lambda}[\slashed{k}_T, \gamma^-] \right)  \ .
  \label{e:Xi_twist_dec}
\end{align}
The amplitudes $A_{1,3}$ and $B_{1,3}$ can be projected out by tracing $\Xi$ multiplied by suitable Dirac matrices.
We report here only the non-zero traces and group these according to the power counting with which they contribute to Eq.~\eqref{e:Xi_twist_dec}: \\
\begin{itemize}

\item
  Twist-2 structures - $O(k^-)$:
  \begin{align}
    \label{e:trace_gm}
    \text{Tr}[\Xi\, \gamma^-] & = 4 A_3 k^- 
  \end{align}

\item
  Twist-3 structures - $O(\Lambda)$:
  \begin{align}
    \label{e:trace_1}
    \text{Tr}[\Xi\, \id] & = 4 \Lambda A_1 \\
    \label{e:trace_gi}
    \text{Tr}[\Xi\, \gamma^i] & = 4 A_3 k_T^i \\
    \label{e:trace_isigmaijg5}
    \text{Tr}[\Xi\, \i \sigma^{ij} \gamma_5] 
                             & = -4\Lambda B_3 \epsilon_T^{ij} 
  \end{align}

\item
  Twist-4 structures - $O(\Lambda^2/k^-)$:
  \begin{align}
    \label{e:trace_gp}
    \text{Tr}[\Xi\, \gamma^+] 
       & = 4 A_3 \frac{k^2 + \vect{k}_T^2}{2k^-} + \frac{4\Lambda^2}{k^-} B_1 \\
    \label{e:trace_isigmaipg5}
    \text{Tr}[\Xi\, \i \sigma^{i+} \gamma_5] 
       & = 4 \frac{\Lambda}{k^-} B_3 \epsilon_T^{ij} {k_T}_j 
  \end{align}

\end{itemize}
In these formulas, the  $\epsilon_T^{ij}=\epsilon^{ij\mu\nu}n_{-\mu}n_{+\nu}$ tensor is the projection of the completely antysimmetric tensor in the transverse plane, and $i,j=1,2$ are transverse Lorentz indices (see Appendix~\ref{a:conv}). 

We note that the trace in Eq.~\eqref{e:trace_gm} corresponds to the inclusive jet function $J$ defined in e.g. Ref.~\cite{Procura:2009vm,Jain:2011xz}. 
As we shall see in Sec.~\ref{ss:spectr_dec}, the amplitudes $A_3$ and $A_1$ are also directly related to the chiral even and chiral odd spectral functions of the quark propagator, respectively.

Moreover, 
note that the Dirac structure associated to the amplitude $B_3$ is time-reversal odd (T-odd)~\cite{Mulders:1995dh,Bacchetta:2000jk,Mulders:2016pln,Accardi:2017pmi}. 
In the correlator that defines parton distribution functions, the T-odd structures are generated by the presence of the gauge link in the transverse plane.  
Since the partonic poles vanish in the correlator that defines fragmentation functions, the T-odd FFs are generated by the interchange of {\em in-} and {\em out-}states induced by a time-reversal transformation rather then by the link structure~\cite{Pijlman:2006vm,Bomhof:2004aw,Boer:2003cm,Meissner:2008yf,Metz:2016swz,Gamberg:2008yt,Gamberg:2010uw}. For this reason the FFs are universal, contrary to TMD PDFs. 
As shown in Section~\ref{ss:sum_operator}, the inclusive jet correlator is related to the fragmentation correlator via an on-shell integration over the hadronic momenta and a sum over all the possible hadronic final states. Since there are no {\em out-}states in Eq.~\eqref{e:invariant_quark_correlator}, we conclude that T-odd structures cannot be present in $\Xi$, namely
\begin{align}
\label{e:B3_zero}
  B_3 = 0 \ .   
\end{align}
This further simplifies Eq.~\eqref{e:Xi_twist_dec} and its Dirac projections in Eq.~\eqref{e:trace_isigmaijg5} and Eq.~\eqref{e:trace_isigmaipg5}. We will briefly return to this point also in the next subsection.

\subsection{Convolution representation and spectral decomposition}
\label{ss:spectr_dec}

We aim now at deriving a spectral representation for the inclusive jet correlator \eqref{e:invariant_quark_correlator}, or, equivalently, for the the gauge invariant quark propagator \eqref{e:invariant_quark_correlator_W}. 

The first step is to rewrite Eq.\eqref{e:invariant_quark_correlator_W} as a convolution of a quark bilinear $\i \widetilde S$ and the Fourier transform $\widetilde W$ of the Wilson line,
\begin{align}
  \label{eq:convolution_Xi}
  \Xi_{ij}(k;n_+) = \text{Disc} \int d^4p\, \colav\,  \lom \i \widetilde S_{ij}(p) \widetilde W(k-p;n_+) \rom \ ,
\end{align}
where 
\begin{align}
\label{e:def_tildeS}
\i \widetilde S_{ij}(p) & = \int \frac{d^4\xi}{(2\pi)^4}\, e^{\i\xi \cdot p}\, \psi_i(\xi) \psibar_j(0)\, , \\
\label{e:def_tildeW}
\widetilde W(k-p;n_+) & = \int \frac{d^4\xi}{(2\pi)^4}\, e^{\i\xi \cdot (k-p)}\, W(0,\xi;n_+) \ .
\end{align}
Note that this convolution representation does not, in itself, depend on the choice of the path for the Wilson line and it is thus generally valid for the study of gauge invariant quark propagators. A careful choice of Wilson line within the generic class discussed in Section~\ref{sss:link_structure} is only needed when relating the gauge invariant propagator to the inclusive jet correlator \eqref{e:invariant_quark_correlator}. 



The convolution representation becomes very useful in combination with the spectral decomposition of the quark bilinear.  The vacuum expectation value of the operator $\i \widetilde S$ is, indeed, the retarded/advanced (according to the sign of $\xi^0$) quark propagator, for which there exist spectral representations~\cite{Bjorken:1965zz}. However, in this paper, we are rather interested in the jet correlator integrated over the sub-dominant $k^+$ component of the quark momentum, and we can, in fact, work with the simpler spectral representation of the Feynman quark propagator. To this end, we introduce the following \textit{\it auxiliary} unintegrated correlator:
\begin{align}
\label{e:Xiprime}
\Xi^\prime_{ij}(k;n_+) = \text{Disc} \int \frac{d^4 \xi}{(2\pi)^4} e^{\i k \cdot \xi} \,
\colav\, \lom\, \timeord \big[ \psi_i(\xi) \psibar_j(0) \big]\, W(0,\xi;n_+) \rom \, ,   
\end{align}
where
only the quark bilinear operator is time ordered.  In Section~\ref{ss:TMD_J_corr}, we will show that under the Wilson line choice discussed in Section~\ref{sss:link_structure}, the $\Xi$ and $\Xi^\prime$ correlators integrated over $k^+$ are, in fact, identical, and we can equivalently work with the latter.

The convolution representation of $\Xi^\prime$ is obtained from Eq.~\eqref{eq:convolution_Xi} by replacing $\i \widetilde S$ with $\i \widetilde S^\prime$, defined as 
\begin{equation}
\label{e:def_tildeSF}
\i \widetilde S^\prime_{ij}(p) = \int \frac{d^4\xi}{(2\pi)^4}\, e^{\i\xi \cdot p}\,
\timeord \big[ \psi_i(\xi) \psibar_j(0) \big] \, .
\end{equation}
This operator can be given a Dirac decomposition assuming invariance under Lorentz and parity transformations~\cite{Bjorken:1965zz}:
\begin{align}
\i \widetilde S^\prime_{ij}(p) = \hat s_3(p^2) \slashed{p}_{ij} + \sqrt{p^2} \hat s_1(p^2) \id_{ij} \ ,
\label{e:quark_bilinear_decomp}
\end{align}
where, for simplicity, we omitted an overall identity matrix in color space, and we can call $\hat{s}_{1,3}$ {\it spectral operators} for reasons that will become clear shortly. 
The correlator $\Xi^\prime$ can then be written as
\begin{align}
\label{e:spectral_convolution}
\Xi^\prime_{ij}(k;n_+) = \text{Disc} \int d^4p\, \colav\, 
\lom 
\Big[\hat s_3(p^2) \slashed{p}_{ij} + \sqrt{p^2} \hat s_1(p^2) \id_{ij} \Big]  
\widetilde W(k-p;n_+)\, 
\rom \ .
\end{align}
%
We can obtain a connection with the K\"allen-Lehman spectral representation of the quark propagator~\cite{Bjorken:1965zz,Weinberg:1995mt,Accardi:2008ne,Accardi:2019luo} by noticing that the Feynman propagator for the quark in momentum space is given by the expectation value of $\i \widetilde S^\prime$ on the interacting vacuum. In turn, the Feynman propagator can be written as a superposition of propagators for (multi)particle states of invariant mass $\mu$~\cite{Bjorken:1965zz,Weinberg:1995mt}:
\begin{equation}
\label{e:Feyn_spec_rep}
\colav\, \lom \i \widetilde S^\prime(p) \rom = \frac{1}{(2\pi)^4} \int_{-\infty}^{+\infty} d\mu^2 
\Big\{ \slashed{p}\, \rho_3(\mu^2) + \sqrt{\mu^2}\, \rho_1(\mu^2) \Big\}\, 
\theta(\mu^2)\,
\frac{\i}{p^2-\mu^2+ \i \epsilon} \ ,
\end{equation}
where the theta function ensures that the spectral functions $\rho_{1,3}$ contribute to the integral only at time-like momenta.\footnote{In Eq.~\eqref{e:Feyn_spec_rep}, there is an extra $(2\pi)^{-4}$ factor with respect to Refs.~\cite{Bjorken:1965zz,Accardi:2008ne,Accardi:2017pmi} because of the normalization of Eq.~\eqref{e:invariant_quark_correlator}, which is customary in the literature dealing with TMD parton distribution and fragmentation functions and the associated non-local operators (see {\em e.g.} Ref.~\cite{Bacchetta:2006tn}).} 
As a consequence of the canonical commutation relations, the spectral function $\rho_{3}$ satisfies~\cite{Zwicky:2016lka,Weinberg:1995mt}
\begin{equation}
  \int_0^{+\infty} d\mu^2 \rho_3(\mu^2) = 1 \ .
\label{e:rho13_positivity_rho3sumrule}
\end{equation}This function can then be interpreted as the probability distribution for a quark to fragment into a multi-particle state of invariant mass $\mu^2$. The $\rho_1$ spectral function does not  satisfy any normalization condition, but, as we will show in Section~\ref{sss:zeta_calc}, is related to the 
mass density of the quark hadronization products. 
Care is, however, needed with these interpretations since the positivity of $\rho_3$ (along with that of $\rho_1$) is not guaranteed in a confined theory. 

Note that, when working in an axial gauge $v \cdot A=0$ as we will do in Section~\ref{s:1h_rules}, we should also add a structure proportional to $\slashed v$ to the decomposition in Eq.~\eqref{e:quark_bilinear_decomp} and to the term in curly brackets in Eq.~\eqref{e:Feyn_spec_rep}, see also Ref.~\cite{Yamagishi:1986bj}. However, in our explicit calculations we will adopt the light-cone gauge, where $v=n_+$, and the additional, gauge-fixing term would only contribute at twist-4 level. Since in this work we limit the applications of our formalism to FF sum rules up to the twist-3 level, for sake of simplicity we have not explicitly written the gauge-fixing term in Eq.~\eqref{e:Feyn_spec_rep}, but we will briefly return on its role in Section~\ref{ss:TMDjet_recap}.


%
%
The discontinuity in Eq.~\eqref{e:spectral_convolution} is completely determined by the discontinuity of  Eq.~\eqref{e:Feyn_spec_rep}. To calculate the latter, we employ the Cutkosky rule~\cite{Cutkosky:1960sp,Bloch:2015efx,Zwicky:2016lka}, by which one simply needs to replace
\begin{equation}
\label{e:Cutkosky_rule}
\frac{1}{p^2-\mu^2 + \i \epsilon} \longrightarrow -2\pi \i\, \delta(p^2 - \mu^2)\, \theta(p^0) 
\end{equation}
at the right hand side of that equation. Namely, as the multiparticle state of invariant mass $\mu^2$ passes the cut, this can be thought of as a set of on-shell particles with positive energy. We thus obtain:
\begin{align}
\nonumber
\text{Disc}\, \colav\, \lom \i \widetilde S^\prime(p) \rom & = \frac{1}{(2\pi)^3} \int_{-\infty}^{+\infty} d\mu^2 
\big\{ \slashed{p}\, \rho_3(\mu^2) + \sqrt{\mu^2}\, \rho_1(\mu^2) \big\}\, 
\theta(\mu^2)\,
\delta(p^2 - \mu^2)\, \theta(p^0) \\
\label{e:cut_Feyn_spec_rep}
& = \frac{1}{(2\pi)^3}\, \big\{ \slashed{p}\, \rho_3(p^2) + \sqrt{p^2}\, \rho_1(p^2) \big\}\, 
\underbrace{\theta(p^2)\, \theta(p^0)}_{=\theta(p^2)\, \theta(p^-)} 
\end{align}
Finally, using the operator decomposition for $\i \widetilde S^\prime$ given in Eq.~\eqref{e:quark_bilinear_decomp}, we obtain the spectral representation for the discontinuity of the expectation values of the operators $\hat{s}_{1,3}$: 
\begin{equation}
\label{e:spec_rep_s13}
(2\pi)^{3}\, \text{Disc}\, \colav\, \lom \hat{s}_{1,3}(p^2) \rom =  \rho_{1,3}(p^2)\, \theta(p^2)\, \theta(p^-) \ .
\end{equation}
It is in this sense, that we can refer to $\hat s_{1,3}$ as spectral operators. 

Note that Eqs.~\eqref{e:cut_Feyn_spec_rep} and~\eqref{e:spec_rep_s13} provide a spectral representation for the quark propagator $\lom \i \widetilde{S}^\prime \rom$ without a Wilson line insertion. It is the purpose of the convolution representation in Eq.~\eqref{e:spectral_convolution}, supplemented by Eq.~\eqref{e:spec_rep_s13} to provide a spectral representation for $\Xi^\prime(k;n_+)$. Its application to the calculation of the $k^+$-integrated jet correlator is discussed in the next Section.


\subsection{The TMD inclusive jet correlator}
\label{ss:TMD_J_corr}

When integrating the inclusive jet correlator over the suppressed $k^+$ quark momentum component, one obtains the TMD inclusive  jet correlator $J$~\cite{Accardi:2019luo},
\begin{align}
\label{e:J_TMDcorr}
  J_{ij}(k^-,\vect{k}_T;n_+)
    & \equiv \frac{1}{2} \int dk^+\, \Xi_{ij}(k;n_+) \nonumber \\
    & = \frac{1}{2}\ \text{Disc} \int \frac{d\xi^+ d^2 \vect{\xi}_T}{(2\pi)^3}\,
      e^{\i k \cdot \xi} \colav
      \lom \psi_i(\xi) \psibar_j(0) W(0,\xi;n_+) 
      \rom_{|_{\xi^-=0}} \ ,
\end{align}
where the $1/2$ normalization factor is justified in Appendix~\ref{a:conv}, and the integrand is now restricted to $\xi^-=0$.

The TMD jet correlator can be decomposed in Dirac structures, with coefficients that can be determined by integrating the projections of $\Xi$ given in Eqs.~\eqref{e:trace_gm}-\eqref{e:trace_isigmaipg5}. 
Following the arguments discussed in Appendix~\ref{a:conv} and using Eq.~\eqref{e:proj_int_J}, we define the projection of $J$ to be:
\begin{equation}
\label{e:def_proj}
J^{[\Gamma]} \ \equiv\ \Tr\bigg[ J\, \frac{\Gamma}{2} \bigg] = 
\frac{1}{2} \int dk^+ \Tr \bigg[ \Xi\, \frac{\Gamma}{2} \bigg] = 
\frac{1}{4} \int dk^+ \Tr \big[ \Xi\, \Gamma \big] \ .
\end{equation}
For twist-2 structures we have:
\begin{align}
\label{e:trace_J_gm}
J^{[\gamma^-]} & =
\frac{1}{2} \int dk^2 A_3(k^2,k^-) 
\ \equiv\ \alpha(k^-) \ .
\end{align}
For twist-3 structures we have:
\begin{align}
\label{e:trace_J_id}
& J^{[\id]} = 
\frac{\Lambda}{2k^-} \int dk^2 A_1(k^2,k^-) 
\ \equiv\ \frac{\Lambda}{k^-} \zeta(k^-)  \\
\label{e:trace_J_gi}
& J^{[\gamma^i]} = 
\frac{k_T^i}{2k^-} \int dk^2 A_3(k^2,k^-) 
= \frac{\Lambda}{k^-} \alpha(k^-) \frac{k_T^i}{\Lambda}  \\
\label{e:trace_J_isigmaijg5}
& J^{[\i \sigma^{ij} \gamma_5]} =
- \frac{\Lambda}{2k^-} \epsilon_T^{ij} \int dk^2 B_3(k^2,k^-) 
\ \equiv\ - \frac{\Lambda}{k^-} \epsilon_T^{ij} \eta(k^-) \ .
\end{align}
For twist-4 structures we have:
\begin{align}
\label{e:trace_J_gp}
J^{[\gamma^+]} & =
\frac{\Lambda^2}{2(k^-)^2} \int dk^2 \bigg[ A_3(k^2,k^-) \frac{k^2+\vect{k}_T^2}{2\Lambda^2} + B_1(k^2,k^-) \bigg] 
\ \equiv\ \frac{\Lambda^2}{(k^-)^2} \omega(k^-,\vect{k}_T^2) 
\end{align}
\begin{align}
\label{e:trace_J_isigmaipg5}
J^{[\i \sigma^{i+} \gamma_5]} & =
\frac{\Lambda^2}{2(k^-)^2} \epsilon_T^{ij} \frac{{k_T}_j}{\Lambda} \int dk^2 B_3(k^2,k^-)
= \frac{\Lambda^2}{(k^-)^2} \epsilon_T^{ij} \frac{{k_T}_j}{\Lambda} \eta(k^-) \ .
\end{align}
Because of the integration over $k^2$, all the functions defined in the previous equations depend only on $k^-$, apart from $\omega$ which has an additional dependence on $\vect{k}_T^2$ that we will discuss in Section~\ref{sss:omega_calc}.
The TMD jet correlator can then be given a twist decomposition in Dirac space as follows: 
\begin{align}
\label{e:J_Dirac}
J(k^-,\vect{k}_T;n_+) & = 
\frac{1}{2} \alpha(k^-) \gamma^+ \\ 
\nonumber
& + \frac{\Lambda}{2k^-} \left[ \zeta(k^-) \id  
											    + \alpha(k^-) \frac{\slashed{k}_T}{\Lambda} 
											    	+ \eta(k^-) \sigma_{\mu\nu} n_-^\mu n_+^\mu \right] \\
\nonumber
& + \frac{\Lambda^2}{2(k^-)^2} \left[ \omega(k^-,\vect{k}_T^2) \gamma^-
														+ \frac{1}{\Lambda} \eta(k^-) \sigma_{\mu\nu} k_T^\mu n_+^\nu \right] \ . 	
\end{align}
Since $B_3 = 0$, according to time-reversal symmetry arguments, we obtain $\eta(k^-) = 0$ and thus the correlator $J$ simplifies to:
\begin{equation}
\label{e:J_Dirac_noeta}
J(k^-,\vect{k}_T;n_+) = 
\frac{1}{2} \alpha(k^-) \gamma^+ 
+ \frac{\Lambda}{2k^-} \left[ \zeta(k^-) \id  
											    + \alpha(k^-) \frac{\slashed{k}_T}{\Lambda} \right] \\
+ \frac{\Lambda^2}{2(k^-)^2} \left[ \omega(k^-,\vect{k}_T^2) \gamma^- \right] \ . 	
\end{equation}
Note that one could also explicitly factor a $\theta(k^-)$ function out of $\alpha$, $\zeta$, and $\omega$. 
The positivity of $k^-$ is indeed guaranteed in any gauge by four-momentum conservation, if one assumes that the particles in the final state all have physical four-momenta. 

The explicit calculation of the coefficients in Eq.~\eqref{e:J_Dirac_noeta} can be carried out with the aid of the convolutional spectral representation discussed in Section~\ref{ss:spectr_dec}. Indeed, recall that in Eq.~\eqref{e:J_TMDcorr} the integrand is restricted to the light front $\xi^-=0$, so that $\xi^2 = -\vect{\xi}_T^2 < 0$ is space-like. Under this condition, the fermion fields anticommute and $ \timeord \big[\psi_i(\xi) \psibar_j(0)\big] = \psi(\xi)\psibar(0)$.
Thus, the integrated version of the correlators $\Xi$ and $\Xi^\prime$ are equivalent,
\begin{equation}
\label{e:J_intXi_intXiprime}
J_{ij}(k^-,\vect{k}_T;n_+) = \frac{1}{2} \int dk^+\, \Xi_{ij}(k;n_+) 
\equiv \frac{1}{2} \int dk^+\, \Xi^\prime_{ij}(k;n_+) \, .
\end{equation}
and one can utilize formulas \eqref{e:spectral_convolution} and \eqref{e:spec_rep_s13} in the calculation of the jet correlator coefficient. This task is carried out in the light-cone gauge in the next three susbsections. 
%

\subsubsection{Calculation of the twist-2 $\alpha$ coefficient}
\label{sss:alpha_calc}

Using the definition of $\alpha$ given in Eq.~\eqref{e:trace_J_gm}, the equivalence Eq.~\eqref{e:J_intXi_intXiprime} between the $\Xi$ and $\Xi^\prime$ integrated correlators, and the convolution representation for $\Xi^\prime$, we find: 
\begin{align}
\label{e:alpha_calc_1}
\alpha(k^-) & 
= \int dk^+\, \text{Disc} \int_{\mathbf M} d^4p\, \colav \lom \hat{s}_3(p^2) p^- \widetilde W(k-p) \rom \\
\nonumber
& = \frac{1}{2}\, \text{Disc} \int_{\mathbf R} dp^2 \int_{\mathbf R} dp^- \colav 
\lom \hat{s}_3(p^2) \int \frac{d\xi^+}{2\pi} e^{\i \xi^+ (k^- - p^-)} 
W_{coll}(\xi^+) \rom \ ,
\end{align}
where the integration domain for $p$ is the whole Minkowski space ($\mathbf M$) and we decompose the integral as $d^4p =  dp^2\, d^2\vect{p}_T\, dp^-/2p^-$. 
From the first to the second line we used Eq.~\eqref{e:def_tildeW} and performed the integrations over $k^+$ and $\vect{p}_T$, fixing $\xi^-=0$ and $\vect{\xi}_T=0$ so that the staple-shaped Wilson line reduces to the straight gauge link in the $n_+$ collinear direction, {\it i.e.}, $W_{coll}(\xi^+)$. Next, we choose the light-cone gauge $A^-=0$ so that the collinear Wilson line reduces to the unity matrix in color space. Finally, performing the integration over $\xi^+$ we obtain:
\begin{align}
\label{e:alpha_calc_2}
\nonumber
\alpha(k^-) & \overset{lcg}{=} \text{Disc} \int_{\mathbf R} dp^2 \int_{\mathbf R} \frac{dp^-}{2} \colav\, \lom \hat{s}_3(p^2) \rom \delta(k^- - p^-) 
= \int_{\mathbf R} dp^2 \int_{\mathbf R} \frac{dp^-}{2} \delta(k^- - p^-) \{ (2\pi)^{-3} \rho_3(p^2) \theta(p^2) \theta(p^-) \} \\
& = \frac{1}{2(2\pi)^3} \bigg\{ \int_0^{+\infty} dp^2 \rho_3(p^2) \bigg\} \theta(k^-) = \frac{\theta(k^-)}{2(2\pi)^3} \ ,
\end{align}
where $lcg$ stresses the use of the light-cone gauge. In the second step we used the representation for the spectral operator $\hat{s}_3$ given in Eq.~\eqref{e:spec_rep_s13}, and in the last one we used the normalization property for $\rho_3$ given in Eq.~\eqref{e:rho13_positivity_rho3sumrule}. We remark that the only dependence on $k^-$ resides in the theta function, and that this result hold, in fact, in any gauge beacuse of the invariance of the jet correlator, hence of the coefficients of its Dirac decomposition. The theta function, which is due to four momentum conservation, and the accompanying numerical coefficients determined by the convention used in the definition of the correlators, also appear in the calculation of the higher-twist $\zeta$ and $\omega$ coefficients to be discussed next.

\subsubsection{Calculation of the twist-3 $\zeta$ coefficient}
\label{sss:zeta_calc}

The $\zeta$ coefficient defined in Eq.~\eqref{e:trace_J_id} is proportional to the trace of the gauge invariant TMD jet correlator $J$. 
Factoring out the $\theta(k^-)$ function, we can thus write
\begin{align}
  \zeta(k^-) = \frac{\theta(k^-)}{2(2\pi)^3\Lambda} M_j \ ,
\end{align}
where $M_j$ is a {\it gauge-invariant} mass term. $M_j$ is in fact independent of $k^-$, and can be interpreted as the inclusive jet's (or the color-averaged dressed quark's) mass, as we will presently show.  

The calculation of $\zeta$ in the light-cone gauge follows closely the procedure outlined in the calculation of $\alpha$. We start from the definition of $\zeta$ given in Eq.~\eqref{e:trace_J_id}, use the convolution representation \eqref{e:spectral_convolution} for the jet correlator, and obtain
\begin{align}
\label{e:zeta_calc_1}
\zeta(k^-) & 
= \frac{k^-}{4\Lambda} \int dk^+\, \text{Disc} \int_{\mathbf M} d^4p\, \colav \lom \sqrt{p^2} \hat{s}_1(p^2) 4 \widetilde W(k-p) \rom \\ 
\nonumber
& = \frac{k^-}{2\Lambda}\, \text{Disc} \int_{\mathbf R} dp^2 \int_{\mathbf R} \frac{dp^-}{p^-} \colav 
\lom \sqrt{p^2} \hat{s}_1(p^2) \int \frac{d\xi^+}{2\pi} e^{\i \xi^+ (k^- - p^-)} W_{coll}(\xi^+) \rom \ , 
\end{align}
where the integrations have been performed as in the case of $\alpha$. In particular the integration over $\vect{p}_T$ has projected the Wilson line on the light cone, leaving us once more with $W_{coll}(\xi^+)$. 
Imposing the light-cone $A \cdot n_+ = 0$ gauge and integrating over $\xi^+$, we obtain:
\begin{align}
\label{e:zeta_calc_2}
\zeta(k^-) & \overset{lcg}{=} \frac{k^-}{2\Lambda} \text{Disc} \int_{\mathbf R} dp^2 \int_{\mathbf R} \frac{dp^-}{p^-} 
\colav\, \lom \sqrt{p^2} \hat{s}_1(p^2) \rom \delta(k^- - p^-) \\
\nonumber
& = \frac{k^-}{2\Lambda} \int_{\mathbf R} dp^2 \int_{\mathbf R} \frac{dp^-}{p^-} \delta(k^- - p^-) \sqrt{p^2} 
\{ (2\pi)^{-3} \rho_1^{}(p^2) \theta(p^2) \theta(p^-) \} 
 = \frac{\theta(k^-)}{2(2\pi)^3\Lambda} \bigg\{ \int_0^{+\infty} dp^2 \sqrt{p^2} \rho_1^{}(p^2) \bigg\} \ ,
\end{align}
where, in going from the first to the second line, we used the representation for the spectral operator $\hat{s}_1$ given in Eq.~\eqref{e:spec_rep_s13}

This calculation shows that the gauge-invariant jet mass $M_j$ has a particularly simple form when choosing the light-cone gauge, being completely determined by the first moment of the ``chiral-odd'' spectral function $\rho_1^{}$:
\begin{equation}
  \label{e:Mj_lcg}
   M_j  \overset{lcg}{=} \int_0^{+\infty} d\mu^2\, \sqrt{\mu^2}\, \rho_1^{}(\mu^2) \ .
\end{equation}
The integral at the right hand side is summing over all the discontinuities of the quark propagator. In this gauge, therefore, $M_j$ can be interpreted as the average mass generated by chirality-flipping processes during the quark's fragmentation, and therefore called ``jet mass'' as proposed in Ref.~\cite{Accardi:2017pmi}. We will elaborate further on this interpretation in Section~\ref{ss:TMDjet_recap}.  
In closing, it is important to remark, that although the explicit dependence of $M_j$ on $\rho_1$ may depend on the choice of gauge,  its numerical value is in fact gauge invariant - and, in particular, independent of $k^-$ as anticipated.

\subsubsection{Calculation of the twist-4 $\omega$ coefficient}
\label{sss:omega_calc}

The calculation of the twist-4 $\omega$ coefficient is more complex than for the $\alpha$ and $\zeta$ coefficients, although the main ideas and techniques discussed in the previous two subsection also apply to this case. 
Since this coefficient appears in Eq.~\eqref{e:J_Dirac_noeta} at twist 4 only, it will not contribute to the fragmentation function sum rules discussed in Section~\ref{s:1h_rules}, that are for now derived up to twist 3. For this reason, we leave a full study of the $\omega$ coefficient for future work, and here we outline its general properties.

From the definition of $\omega$ in Eq.~\eqref{e:trace_J_gp} and using the convolution representation in Eq.~\eqref{e:spectral_convolution} the $\omega$ coefficient reads:
\begin{align}
\label{e:omega_calc_1}
\nonumber 
\omega(k^-,\vect{k}_T^2) & = \bigg(\frac{k^-}{\Lambda}\bigg)^2 
\int dk^+\, \text{Disc} \int_{\mathbf M} d^4p\, \colav \lom \hat{s}_3(p^2)
\frac{p^2+\vect{p}_T^2}{2p^-} \widetilde W(k-p) \rom\\ 
   & \equiv\ \llangle \frac{p^2}{(p^-)^2} \rrangle
      + \llangle \frac{\vect{p}_T^2}{(p^-)^2} \rrangle \ .
\end{align}



The integral involving $p^2$ in Eq.\eqref{e:omega_calc_1} can be calculated following the same procedure used for the $\zeta$ coefficient. One obtains
\begin{equation}
\label{e:omega_avp2}
  \llangle \frac{p^2}{(p^-)^2} \rrangle = \frac{\theta(k^-)}{4 \Lambda^2 (2\pi)^3} \, \mu_j^2 \ ,
\end{equation}
where the $\theta(k^-)$ function arises as for the $\alpha$ and $\zeta$ coefficients, and (similarly to $M_j$) $\mu_j^2$ has a particularly simple form in the light-cone gauge:
\begin{equation}
  \label{e:Kj2_lcg} 
   \mu_j^2 \overset{lcg}{=} \int_0^{+\infty} d\mu^2\, \mu^2\, \rho_3^{}(\mu^2) \ .
\end{equation}Unlike $M_j$, however, this is not gauge-invariant.  Given the properties of $\rho_3$ in Eq.~\eqref{e:rho13_positivity_rho3sumrule}, $\mu_j^2$ can be interpreted as the average invariant mass squared directly generated by the quark as it fragments into the final state\footnote{One has to be careful, though, with this interpretation since the positivity of $\rho_3$ is not guaranteed in a confined theory.}. 

The calculation of the $\llangle \vect{p}_T^2 / (p^-)^2 \rrangle$ term is more involved because one cannot immediately integrate over $\vect{p}_T$ and project the integrand on the light cone. To achieve that, one needs first to remove the explicit dependence of the integrand on $\vect{p}_T^2$ using
\begin{equation}
\label{e:pT2_deriv}
  \vect{p}_T^2\, e^{\i \vect{\xi}_T \cdot (\vect{p}_T - \vect{k}_T)}
  = \bigg( -\frac{\partial}{\partial \vect{\xi}_T^\alpha}
    \frac{\partial}{\partial {\vect{\xi}_T}_\alpha} 
    -2\i \vect{k}_T^\alpha \frac{\partial}{\partial \vect{\xi}_T^\alpha} 
    + \vect{k}_T^2 \bigg)\, e^{\i \vect{\xi}_T \cdot (\vect{p}_T - \vect{k}_T)} \ .
\end{equation}
One then obtains
\begin{align}
\label{e:omega_avpT2}
  \llangle \frac{\vect{p}_T^2}{(p^-)^2} \rrangle 
    = \frac{\theta(k^-)}{4 \Lambda^2 (2\pi)^3} \,
    \Big( \vect{k}_T^2 + \tau_j^2 \Big) \ ,
\end{align}
where 
\begin{align}
\label{e:Tjdef}
\theta(k^-) \, \tau_j^2 = (2\pi)^3 (k^-)^2\, \text{Disc} 
  & \int_{\mathbf M} d^4 p\, \colav \frac{1}{(p^-)^2} \lom \hat{s}_3(p^2) 
  \int \frac{d\xi^+ d^2\vect{\xi}_T}{(2\pi)^3} e^{\i \xi^+ (k^- - p^-)} \\
\nonumber
& \times \bigg( -\frac{\partial}{\partial \vect{\xi}_T^\alpha} \frac{\partial}{\partial {\vect{\xi}_T}_\alpha} 
-2\i \vect{k}_T^\alpha \frac{\partial}{\partial \vect{\xi}_T^\alpha} \bigg) 
e^{\i \xi_T \cdot (\vect{p}_T - \vect{k}_T)} W_{TMD}(\xi^+, \xi_T) \rom \ .
\end{align}
The $\vect{k}_T^2$ term in Eq.\eqref{e:omega_avpT2} is a purely kinematical effect of the initial parton's  non-zero transverse momentum. The $\tau_j^2$ term can be interpreted 
as the average squared transverse momentum of the fragmented hadrons relative to the quark axis. This quantity also characterizes the jet's transverse shape: the larger $\tau_j^2$ the less aligned the final state is to the initial quark; in this sense, we can call $\tau_j^2$ the ``jet broadening" parameter.

\subsection{Summary and interpretation of $\boldmath M_j$}
\label{ss:TMDjet_recap}

Inserting the expressions of the $\alpha$, $\zeta$, $\omega$ coefficients in Eq.~\eqref{e:J_Dirac_noeta} we obtain the following decomposition for the TMD jet correlator:
\begin{align}
\label{e:J_Dirac_explicit}
  J(k^-,\vect{k}_T;n_+)
    = \frac{\theta(k^-)}{4(2\pi)^3\, k^-} \, 
    \bigg\{ k^-\, \gamma^+ + \slashed{k}_T + M_j \id + \frac{K_j^2 + \vect{k}_T^2}{2k^-} \gamma^- \bigg\} \ ,
\end{align}
with the ``jet virtuality''
\begin{align}
  K_j^2 = {\mu_j^2 + \tau_j^2 + \textit{g.f.t.}}
\end{align}
receiving contributions from the invariant mass directly produced in the quark fragmentation process ($\mu_j^2$, Eq.\eqref{e:Kj2_lcg}), from the final state jet broadening ($\tau_j^2$, Eq.\eqref{e:Tjdef}), and from a gauge fixing term [$g.f.t.$]. The latter symbolically represents the potential contributions from a structure proportional to $\slashed{v} = \slashed{n}_+$ in Eq.~\eqref{e:quark_bilinear_decomp} and~\eqref{e:Feyn_spec_rep}, that, as discussed, we do not consider further in this article. As it happens to the jet mass $M_j$~\eqref{e:Mj_lcg} the jet virtuality $K_j^2$ is also a gauge invariant quantity. 

The expression in brackets in Eq.~\eqref{e:J_Dirac_explicit} generalizes the familiar term appearing in the numerator of the free quark propagator:
\begin{align}
  \slashed{k} + m = k^-\, \gamma^+ + \slashed{k}_T + m \id
    +  \frac{m^2 + \vect{k}_T^2}{2k^-}\, \gamma^- \ . 
\end{align}We can see that the current quark mass generalizes to the jet mass, $m \rightsquigarrow M_j$, and the mass shell generalizes to the jet's virtuality, $m^2 \rightsquigarrow K_j^2$. Conversely, using the non-interacting propagator's spectral functions $\rho_{1,3} \propto \delta(\mu^2-m^2)$ in Eqs.~\eqref{e:Kj2_lcg}, we obtain $M_j=m$; furthermore, neglecting the contribution of the Wilson line and the gauge fixing term, we obtain $K_j^2=m^2$. 

Overall, the jet correlator~\eqref{e:J_Dirac_explicit} can be thought as a propagating particle of mass $M_j$, that is however off the mass shell because its virtuality $K_j^2 = \mu_j^2  + \tau_j^2 + \textit{g.f.t.}$ is in general different from $M_j^2$. However, it may be dangerous to push this interpretation beyond the kinematic level, because the jet mass cannot necessarily be interpreted as a pole mass.
%
In fact, let us consider the non-perturbative Feynman quark propagator in momentum space expressed in terms of a renormalization factor $Z(p^2)$ and a mass function $M(p^2)$~\cite{Roberts:2007jh,Roberts:2015lja,Siringo:2016jrc,Zwicky:2016lka,Solis:2019fzm})\footnote{As for the spectral representation in Eq.~\eqref{e:Feyn_spec_rep}, there is an additional $1/(2\pi)^4$ factor with respect to the expression given in e.g. Refs.~\cite{Roberts:2007jh,Roberts:2015lja,Siringo:2016jrc,Zwicky:2016lka,Solis:2019fzm} in order to match the convention for the Fourier transform used in Eq.~\eqref{e:invariant_quark_correlator}.}:
\begin{equation}
\label{e:SF_mass}    
\i S_F(p) =  \frac{\i Z(p^2)}{(2\pi)^4 [\slashed{p} - M(p^2)]} \, . 
\end{equation}
By comparing this expression with the spectral representation for the Feynman propagator presented in Eq.~\eqref{e:Feyn_spec_rep} with $p^0>0$ and using the definition~\eqref{e:Mj_lcg} of $M_j$ in the light-cone gauge, we find that
\begin{align}
\label{e:relation_Mj_M}
M_j \overset{lcg}{=} \int_{-\infty}^{+\infty} dp^2\, \theta(p^2)\, \sqrt{p^2}\, \rho_1(p^2) = 
\frac{\i}{2\pi} \int_{-\infty}^{+\infty} dp^2\, \text{Disc}\, \frac{Z(p^2) M(p^2)}{p^2 - M^2(p^2)} \, . 
\end{align}
This equation relates the gauge-invariant and scale-dependent jet mass $M_j$ in the light-cone gauge and the gauge-dependent and scale-invariant mass function $M(p^2)$.  
The scale dependence of the jet mass is provided by the (implicit) scale dependence of the renormalized spectral function $\rho_1$ on the one hand, and on the other hand is accounted for by the $Z(p^2)$ renormalization function~\cite{Roberts:2015lja,Roberts:2007jh,Solis:2019fzm}. 
If the propagator $\i S_F$ in Eq.~\eqref{e:SF_mass} had a single pole 
at $p^2=M^2(p^2)\equiv M^2_p$
and no branch cut, 
by a simple application of Cutkosky's rule one would obtain $M_j = Z(M_p^2)\, M_p$ -- i.e., the jet mass could be identified with the renormalized pole mass. In general, however, $M_j$ is summing all the discontinuities of the non-perturbative propagator, hence also over the mass spectrum continuum, and can be different from zero even if no pole, in fact, exists. A more universal interpretation of $M_j$ can be obtained by considering the jet correlator as represented in Eq.~\eqref{e:invariant_quark_correlator}. $M_j$ can then be thought as the sum over the masses of all physical states overlapping with a quark, weighted by the amplitude squared of that particular quark to multi-hadron state transition.


From a heuristic point of view, 
one can think 
of $M_j$ as a gauge-invariant mass scale that characterizes the physics of a {\em color-averaged} (or {\em color-screened}) dressed quark. 
In light of this interpretation, it is possible to subtract from $M_j$ the current quark mass component responsible for the explicit breaking of the chiral symmetry, and isolate a dynamical component generated by quark-gluon interactions and responsible for the dynamical breaking of the chiral symmetry. 
This suggest the decomposition
\begin{equation}
\label{e:Mj_decomp}
M_j = m + m^{\text{corr}} \, ,
\end{equation} 
where $m$ is the current quark mass, $m^{\text{corr}}$ is the dynamical mass, and all terms have an implicit renormalization scale dependence. In perturbation theory $m^{\text{corr}}_{\text{pert}} \propto m$ vanishes in the chiral $m \rightarrow 0$ limit. Thus the dynamical mass $m^{\text{corr}} = M_j - m$ can also be thought as an order parameter for dynamical chiral symmetry breaking. 
As will be discussed in details in Section~\ref{s:1h_rules}, the decomposition~\eqref{e:Mj_decomp} is also particularly meaningful in light of the equations of motion that relate the twist-2 and twist-3 fragmentation functions, and we will see that this mass is quantitatively related to quark-gluon-quark correlations. For this reason in the following we will refer to it as the {\em correlation mass}. 

Crucially, all this discussion is not merely of theoretical interest
because, in fact $M_j$ and $m^{\text{corr}}$ can couple to the target's transversity PDF in inclusive DIS processes \cite{Accardi:2008ne},
and can furthermore be related to the chiral-odd twist-3 fragmentation functions $E^h$ and $\widetilde{E}^h$ by momentum sum rules measurable in semi-inclusive processes~\cite{Accardi:2017pmi,Accardi:2019luo}. 
We are therefore offered the possibility of comparing calculations of the quark spectral functions, nowadays directly possible in Minkowski space~\cite{Solis:2019fzm,Siringo:2016jrc}, to experimentally measurable quantities: this is a non-trivial feature in the case of particles such as quarks that do not appear in the physical spectrum of the theory because of color confinement. 

In summary, when investigating the transition of a quark propagating in the QCD vacuum into a set of detectable hadrons in terms of the higher-twist components of the jet correlator~\eqref{e:J_Dirac_explicit}, one is provided with a with a rather concrete window on color confinement and the dynamical generation of mass. We will revisit these points in Section~\ref{sss:DCSB}, after an in-depth discussion of the connection of the inclusive jet correlator with the single-inclusive fragmentation correlator, and the ensuing fragmentation function sum rules.


\section{Momentum sum rules for single-hadron fragmentation functions}
\label{s:1h_rules}

In this section, we will establish a sum rule at the correlator level between the single-hadron fragmentation correlator and the inclusive jet correlator, and systematically exploit this to derive explicit sum rules for fragmentation functions up to the twist-3 level. We will recover known sum rules, and derive a number of new ones. As we will discuss, the interest of these sum rules also extends beyond their application to phenomenological fits.

\subsection{The single-hadron fragmentation correlator}
\label{ss:1h_inclusive_FF}

The unintegrated correlator describing the fragmentation of a quark into a single hadron (or ``single-inclusive" correlator) is defined as~\cite{Mulders:1995dh,Goeke:2003az,Bacchetta:2006tn,Meissner:2007rx,Mulders:2016pln,Metz:2016swz,Echevarria:2016scs}
\begin{align}
  \label{e:1hDelta_corr}
  & \Delta^h_{ij}(k,P,S) = \sum_X \int \frac{d^4 \xi}{(2\pi)^4} e^{\i k \cdot \xi} \colav \lom 
  \timeord \big[ W_1(\infty,\xi) \psi_i(\xi) \big] \, 
  |P S X \rangle 
  \langle P S X| \, 
  \antitimeord \big[ \overline{\psi}_j(0) W_2(0,\infty) \big] \rom\, ,
\end{align}
where $k$ is the quark's four-momentum, $h$ is an identified hadron with four-momentum $P$ and spin $S$, and $X$ represents the quantum numbers of all unobserved hadrons in the final state. 
The vector $S^\mu$ is the covariant spin vector associated to the Bloch representation for a hadron with spin $1/2$~\cite{Bacchetta:2006tn,Mulders:1995dh}. 
The remarks given in Section~\ref{s:jetcor} about the importance of the color average for the definition of the inclusive jet correlator also apply to the fragmentation correlator~\eqref{e:1hDelta_corr}. A diagrammatic interpretation is given in Figure~\ref{f:cut_diagrams}(b).

In the following we will deal only with the correlator describing the fragmentation of a quark into an unpolarized or spinless hadron $\Delta^h(k,P)$, which is defined as a sum of the polarization-dependent correlator over the polarization of the identified hadron:
\begin{align}
  \label{e:1hDelta_corr_unpol}
  & \Delta^h_{ij}(k,P) = \sum_{S} \Delta^h_{ij}(k,P,S) \ .
\end{align}Letting the sum over $S$ act on the right hand side of Eq.~\eqref{e:1hDelta_corr}, one obtains an explicit definition by substituting $|PSX\rangle \langle PSX|$ with $|PX\rangle \langle PX|$ in that equation.

Let us now focus on the structure of the $| P X \rangle$ final state. This is composed of one identified hadron $h$ with momentum $P$ and a remnant $X$. Following the approach of Ref.~\cite{Levelt:1993ac}, we assume that:
\begin{equation}
\label{e:completeness_X}
\sum_X | X \rangle \langle X | = \id = \sum_{n=0}^{+\infty} \id_n \ .
\end{equation}
The first equality in Eq.~\eqref{e:completeness_X} is the completeness relation for the {$|X\rangle$} states, {\it i.e.}, the resolution of the identity in terms of the projectors $| X \rangle \langle X |$. The second equality decomposes the identity as a sum of identity operators $\id_n$ acting in the sub-space spanned by $n$--hadron states. These can be explicitly represented as 
\begin{equation}
\label{e:def_1n}
\id_n = \frac{1}{n!} \int d\widetilde{K}_1 \cdots d\widetilde{K}_n 
a^\dagger(K_1) \cdots a^\dagger(K_n) 
\rom \lom
a(K_1) \cdots a(K_n) \ ,
\end{equation}
where, for ease of notation we combined the momentum $K_i$ and flavor $h_i$ of the $i$-th unobserved hadron into a single $\widetilde K_i$ variable, and $a(\widetilde K_i)$, $a^\dagger(\widetilde K_i)$ are the associated annihilation and creation operators. The integration reads $\int d\widetilde K_i \equiv \sum_{h_i} \int d^3K_i / [(2\pi)^3 2 E_i]$, and as before a sum over the hadron spin is understood when the corresponding index is not explicitly written.
Using Eqs.~\eqref{e:completeness_X} and \eqref{e:def_1n}, we can recast the sum over the projectors $|PX\rangle \langle PX|$ as:
\begin{align}
\nonumber
\sum_X | P X \rangle \langle P X | & = | P \rangle \langle P | 
+ \int d\widetilde K_1 a^\dagger(\widetilde K_1) | P \rangle \langle P | a(\widetilde K_1) 
+ \frac{1}{2} \int d\widetilde K_1 d\widetilde K_2 a^\dagger(\widetilde K_1) a^\dagger(\widetilde K_2) | P \rangle \langle P | a(\widetilde K_1) a(\widetilde K_2) + \cdots \\
\label{e:1h_proj_to_op} &  = a_h^\dagger \bigg(  \sum_{n=0}^{+\infty} \id_n  \bigg) a_h = a_h^\dagger a_h \ , 
\end{align}
where we have used $a^\dagger(\widetilde K_i) \rom = | \widetilde K_i \rangle$, and $a_h^\dagger \rom = |P \rangle$ creates the identified hadron $h$ from the vacuum.
Using Eq.~\eqref{e:1h_proj_to_op}, we obtain 
\begin{equation}
\label{e:1hDelta_aa}
\Delta^h_{ij}(k,P) = \int \frac{d^4 \xi}{(2\pi)^4} e^{\i k \cdot \xi} \colav \lom 
\timeord \big[ W_1(\infty,\xi) \psi_i(\xi) \big] \, 
(a_h^\dagger a_h) \,
\antitimeord \big[ \overline{\psi}_j(0) W_2(0,\infty) \big] \, 
\rom \ , 
\end{equation}
where it is understood that $a_h = a_h(P,S)$ and the same for $a_h^\dagger$. 
For brevity of notation, in the following we work with the same Wilson lines $W_{1,2}$ we have chosen for the inclusive jet correlator $\Xi$ and drop the (anti)time-ordering operators.
%
The expansion of this correlator on a basis of Dirac structures can be obtained from the one given in Refs.~\cite{Bacchetta:2004zf,Goeke:2005hb} for the distribution correlator by replacing the target hadron momentum with the produced hadron momentum, the target mass with the produced hadron mass, by interchanging $n_-$ with $n_+$, and neglecting the structures related to the polarization of the produced hadron:  
\begin{align}
\label{e:1hDelta_ampl}
\Delta^h(k, P) 
&= M_h A_1 \id + A_2 \slashed{P} + A_3 \slashed{k} + \frac{A_4}{M_h} \sigma_{\mu\nu} P^\mu k^\nu + \\
\nonumber & + \frac{M_h^2}{P \cdot n_+} B_1 \slashed{n}_+ + 
\frac{M_h}{P \cdot n_+} B_2 \sigma_{\mu\nu} P^\mu {n_+^\nu} + 
\frac{M_h}{P \cdot n_+} B_3 \sigma_{\mu\nu} k^\mu n_+^\nu +
{\frac{1}{P \cdot n_+}\, B_4\, \epsilon_{\mu\nu\rho\sigma}\, \gamma^\mu\, \gamma_5\, P^\nu\, k^\rho\, n_+^\sigma} \ .
\end{align}The amplitudes $A_i$ and $B_i$ are functions of the Lorentz scalars $k \cdot n_+$, $k \cdot P$, $k^2$. The terms proportional to $n_+$ originate from the path defining the Wilson lines $W_{1,2}$, that provides one additional vector beside $k$ and $P$ with which to carry out the decomposition. These terms generate TMD and collinear structures that appear only at subleading twist~\cite{Mulders:1995dh,Mulders:2016pln,Bacchetta:2006tn}. 
In keeping with the conventions of \cite{Bacchetta:2006tn}, we have introduced a power-counting scale $M_h$ equal to the mass of the identified hadron. This choice is not mandatory, and only affects the normalization of the above defined amplitudes and of the related fragmentation functions to be introduced below. For example, a flavor-independent choice of scale, such as $\Lambda$ used for the jet expansion of the inclusive jet correlator in Eq.~\eqref{e:Xi_twist_dec},  would slightly simplify a number of the sum rules to be discussed later. However the present choice of $M_h$ not only agrees with most of the literature on TMD FFs, but also suggests interesting physical interpretations for these sum rules. 
It is also of interest to note that, formally, Eq.~\eqref{e:jet_ampl} for the decomposition of the inclusive correlator $\Xi(k;n_+)$ can be obtained from Eq.~\eqref{e:1hDelta_ampl} by replacing the hadron four-momentum $P$ with the parton four-momentum $k$, and by replacing the hadron mass $M_h$ with the power counting scale $\Lambda$.

The fragmentation process can be studied either in the parton or in the hadron frame~\cite{Collins:2011zzd,Levelt:1993ac}. The Lorentz transformation between these and its consequences are discussed in detail in Appendix~\ref{a:frame_dep}\footnote{See also Ref.~\cite{Mulders:2016pln} and Section~12.4.1 in Ref.~\cite{Collins:2011zzd}.}. 
In the parton frame, defined such that the parton's transverse momentum $\vect{k}_T=0$, one can interpret the fragmentation correlator as the probability density for the quark to fragment into a hadron of a given flavor $h$ and momentum $P$, with $\vect{P}_T$ generically non zero \cite{Collins:2011zzd,Metz:2016swz}. 
The parton frame, however, turns out not to be convenient in derivations of factorization theorems and calculations of semi-inclusive cross sections: the partonic momenta are integrated over, and the parton frame axes are not fixed (see Chap. 12 in Ref.~\cite{Collins:2011zzd}). In this case, it is preferable to utilize the hadron frame, where the experimentally observable hadron's 3-momentum determines the $z$ direction, so that $\vect{P}_T=0$. In this frame, it is the quark's transverse momentum that, in general, has a non-zero value. 
%

Since we are not dealing with a specific scattering process and we want to connect the fragmentation correlator to the invariant quark propagator, we choose to work from now on in the parton frame. 
Namely, we consider a base in Minkowski space composed of the light-cone $n_+$ and $n_-$ vectors such that $k=k^+ n_+ + k^- n_-$, and two transverse four-vectors $n_1$ and $n_2$. 
This basis not only determines the coordinates of any four-vector under consideration, but will also be used to define a set of parton-frame TMD fragmentation functions, as we will discuss next.


In calculations of semi-inclusive hadron production cross sections one deals with the fragmentation correlator integrated over the subdominant quark momentum component $k^+$. 
According to the convention outlined in Appendix~\ref{a:conv} this is defined as 
\begin{align}
\label{eq:integratedFFcorr}
\Delta^h(z,P_T) \equiv \frac{1}{2z} \int
dk^+ \, \Delta^h(k, P)_{k^- = P^-/z } \ ,
\end{align} 
which corresponds to:
\begin{equation}
\label{e:1hDelta_TMDcorr}
\Delta^h_{ij}(z,P_T) = \frac{1}{2z} \int \frac{d \xi^+ d^2 \vect{\xi}_T}{(2\pi)^3} e^{\i k^- \xi^+} 
\colav\, \text{Disc}\, \lom 
W_1(\infty,\xi) \psi_i(\xi) (a_h^\dagger a_h) \overline{\psi}_j(0) W_2(0,\infty)  
\rom_{\begin{subarray}{l} \xi^-=0 \\ k^- = P^-/z \end{subarray}}  \ . 
\end{equation}
In the parton frame, this can be expanded in Dirac structures and parametrized in terms of TMD FFs up to twist 3 as: 
\begin{align}
\label{e:1hDelta_TMDcorr_param}
\Delta^h(z,P_T) & = 
\frac{1}{2} \slashed{n}_- D^h_1(z,P_T^2)  
- \i \frac{ \big[ \slashed{P}_T, \slashed{n}_- \big]}{4 z M_h} H_1^{\perp\, h}(z,P_T^2)
+ \frac{M_h}{2 P^-} E^h(z,P_T^2) 
\\
\nonumber & 
- \frac{\slashed{P}_T}{2 z P^-} D^{\perp\, h}(z,P_T^2) 
+ \frac{\i M_h}{4 P^-} \big[ \slashed{n}_-, \slashed{n}_+ \big] H^h(z,P_T^2)
- \frac{1}{2 z P^-} \gamma_5 \epsilon_T^{\rho\sigma} \gamma_\rho {P_T}_\sigma G^{\perp\, h}(z,P_T^2) \ ,
\end{align}
where the Dirac structures and the TMDs explicitly depend on the hadron momentum $P_T$. It is important to note that this decomposition of the TMD fragmentation correlator depends on the choice of the light cone basis vectors, in our case the parton-frame basis discussed above, even though for simplicity of notation no explicit index is introduced to remind us of this fact.
In Refs.~\cite{Mulders:1995dh,Bacchetta:2006tn,Metz:2016swz} an analogous decomposition is given, instead, in the hadron frame, as is standard procedure in the TMD literature. 
The relation between the hadron- and parton-frame decomposition, and therefore between the hadron- and parton-frame TMD fragmentation functions, is discussed in detail in Appendix~\ref{a:frame_dep}. 

Eq.~\eqref{e:1hDelta_TMDcorr_param} can be also re-arranged as a sum of terms with definite rank in $P_T$~\cite{Boer:2016xqr}:
\begin{equation}
\label{e:1hDelta_r0_r1}
\Delta^h(z,P_T) = \Delta_0^h(z,P_T^2) + {P_T}_\alpha\, \Delta_1^{h\, \alpha}(z,P_T^2) \, , 
\end{equation}
where
\begin{align}
\label{e:1hDelta_r0}
& \Delta_0^h(z,P_T^2) = 
\frac{1}{2} \slashed{n}_- D^h_1(z,P_T^2) + 
\frac{M_h}{2 P^-} E^h(z,P_T^2) + 
\frac{\i M_h}{4 P^-} \big[ \slashed{n}_-, \slashed{n}_+ \big] H^h(z,P_T^2) \, , \\
\label{e:1hDelta_r1}
& \Delta_1^{h\, \alpha}(z,P_T^2) = 
- \i \frac{ \big[ \gamma_T^\alpha, \slashed{n}_- \big]}{4 z M_h} H_1^{\perp\, h}(z,P_T^2)
- \frac{\gamma_T^\alpha}{2 z P^-} D^{\perp\, h}(z,P_T^2) 
- \frac{1}{2 z P^-} \gamma_5 \epsilon_T^{\rho\alpha} \gamma_\rho G^{\perp\, h}(z,P_T^2) \, .
\end{align}
The subscript $0,1$ refers to the rank in $P_T$ of the associated structures~\cite{Boer:2016xqr} and, in the following, we will collectively refer to $D_1^h$, $E^h$, $H^h$ as the rank 0 TMD FFs, and to $H_1^{\perp\, h}$, $D^{\perp\, h}$, $G^{\perp\, h}$ as the rank 1 TMD FFs. 
From Eq.~\eqref{e:1hDelta_r0_r1} one can see that~\cite{Bacchetta:2006tn}
\begin{equation}
\label{e:1hDelta_integrated}
\Delta^h(z) = \int d^2 \vect{P}_T\, \Delta^h(z,P_T) = \int d^2 \vect{P}_T\, \Delta_0^h(z,P_T) \, , 
\end{equation}
where the collinear fragmentation correlator $\Delta^h(z)$ is defined as
\begin{equation}
\label{e:1hDelta_coll_op}
\Delta^h_{ij}(z) = \frac{z}{2} \int \frac{d \xi^+ }{2\pi} e^{\i \xi^+ P^-/z} 
\colav\, \text{Disc}\, \lom 
W_1(\infty,\xi) \psi_i(\xi) (a_h^\dagger a_h) \overline{\psi}_j(0) W_2(0,\infty)  
\rom_{\begin{subarray}{l} \xi^-=\vect{\xi}_T=0 \\ P_T=0 
\end{subarray}}  \ . 
\end{equation}
%
Moreover, $\Delta(z)$ can be parametrized as~\cite{Bacchetta:2006tn}:
\begin{equation}
\label{e:1hDelta_coll}
\Delta^h(z) = 
\frac{1}{2} \slashed{n}_- D^h_1(z) + 
\frac{M_h}{2 P^-} E^h(z) + 
\frac{\i M_h}{4 P^-} \big[ \slashed{n}_-, \slashed{n}_+ \big] H^h(z) \, .
\end{equation}
The rank-1 term $\Delta_1^{h\, \alpha}(z,P_T^2)$ in Eq.~\eqref{e:1hDelta_r0_r1} does not contribute at the collinear level~\eqref{e:1hDelta_integrated} since the explicit ${P_T}_\alpha$ factor sets the associated integral over $\vect{P}_T$ to zero. 
For this reason, we will only be able to derive constraints on the integral of rank-0 TMD FFs, but not of rank-1 FFs (see Section~\ref{ss:mu_transverse}). On the contrary, by weighting $\Delta^h(z,P_T)$ by $P_T^\alpha$ and integrating over the transverse momentum one can obtain sum rules for the first moment of rank-1 TMD FFs. 
Note that Eq.~\eqref{e:1hDelta_integrated}, relating the collinear correlator to the TMD correlator integrated over $\vect{P}_T$, is only valid for bare, {\it i.e.}, non-renormalized fields in perturbative QCD~\cite{Collins:2011zzd}.\footnote{The identification of an integrated TMD FF with the corresponding collinear FF is also valid in certain models of QCD, where the integration over the transverse momentum can be regularized introducing a phenomenological scale to suppress the large momentum region. An example is the parton model in Gaussian approximation~\cite{Signori:2013mda,Anselmino:2013lza}. Other examples include the spectator diquark model~\cite{Bacchetta:2008af} and the Nambu--Jona-Lasinio model~\cite{Matevosyan:2011vj}.} 
The same is also true when one identifies the integrated TMD FFs $f = D_1^h, E^h, H^h$ with their collinear counterparts in Eq.~\eqref{e:1hDelta_coll}, \textit{i.e.}, takes $f(z) \ \equiv\ \int d^2 \vect{P}_T f(z,P_T^2)$.

\subsection{Connection between the fragmentation and jet correlators}
\label{ss:sum_operator}

We can now discuss a momentum sum rule connecting the unintegrated single-hadron fragmentation correlator to the inclusive jet correlator.
We work in the context of field theory, taking inspiration from, but generalizing, the strategy outlined in Ref.~\cite{Meissner:2010cc}.
In particular, the authors of that reference directly manipulate the $k^+$-integrated TMD correlator, withouth the notion of the jet correlator, and limit their attention to a restricted number of Dirac structures.
Instead, we prove the sum rule at the level of unintegrated correlators, then specialize this to the TMD correlators, and from there derive the sum rules for the fragmentation functions. As a result, we are able to extend the formalism to include all twist-2 and twist-3 FFs. Let us also stress from the outset that, as discussed in Ref.~\cite{Meissner:2010cc}, the proof is only valid for unpolarized correlators and FFs. In the polarized case one would just obtain trivial identities.
The methods utilized here have also  been used in Ref.~\cite{Anselmino:2011ss} to prove momentum sum rules for the quark fracture functions and reduce these to parton distribution functions,  but without considering Wilson line insertions as we do, instead, in this paper. 

Our starting point are the definitions of the unintegrated correlators $\Xi$ and $\Delta^h$ in Eq.~\eqref{e:invariant_quark_correlator} and Eq.~\eqref{e:1hDelta_corr}, respectively. 
Let us then consider the following quantity:
\begin{equation}
\label{e:average_Ph}
\sum_h \sum_{S} 
\int \frac{d^4 P}{(2\pi)^4}\, (2\pi) \delta(P^2 - M_h^2) 
P^\mu \Delta^h(k,P,S)  \, .
\end{equation}
The integration is performed in Minkowski space over the on-shell momentum $P$ of the detected hadron with mass $M_h$, and the sum extends to all hadron spin states and species. Eq.~\eqref{e:average_Ph} can be loosely understood as providing one with the average four-momentum of the produced hadrons, if one considers $\Delta$ as a probability distribution in hadron momentum and spin. This interpretation becomes explicit in the parton frame for the $\gamma^+$ projection of the $k^+$-integrated $\Delta$ correlator \cite{Mulders:2016pln}. 

Let us now introduce the $\hat {\vect P}_h^\mu$ hadronic momentum operator associated to the vector $P^\mu$ of the identified hadron in the framework of second quantization~\cite{Weinberg:1995mt,Collins:1981uw}:
\begin{equation}
\label{e:Ph_op_1}
\hat{\vect{P}}_h^\mu = \sum_{S} \int \frac{dP^- d^2 \vect{P}_T}{2P^- (2\pi)^3} P^\mu\, \hat{a}_h^\dagger(P,S) \hat{a}_h(P,S) 
\end{equation}
as well as the inclusive $\hat{\vect P}^\mu$ momentum operator, that also appears in Eq.~(4.25) of Ref.~\cite{Collins:1981uw} and in Ref.~\cite{Meissner:2010cc}:
\begin{equation}
\label{e:P_op_1}
\hat{\vect{P}}^\mu = \sum_h \hat{\vect{P}}_h^\mu \ . 
\end{equation}
Using Eqs.~\eqref{e:Ph_op_1} and \eqref{e:P_op_1}, the average four-momentum defined in Eq.~\eqref{e:average_Ph} can be further manipulated:
\begin{align}
\label{e:unintegrated_sum_rule}
\nonumber
\sum_{h\, S} 
\int \frac{d^4 P}{(2\pi)^4}\, (2\pi) \delta(P^2 - M_h^2) 
P^\mu \Delta^h(k,P,S)  & = 
\sum_{h\, S} 
\int \frac{dP^- d^2 \vect{P}_T}{(2\pi)^3 2P^-} P^\mu 
\int \frac{d^4 \xi}{(2\pi)^4} e^{\i k \cdot \xi} 
\lom W_1 \psi_i(\xi)(a_h^\dagger a_h) \overline{\psi}_j(0) W_2 \rom \\
& = \int \frac{d^4 \xi}{(2\pi)^4} e^{\i k \cdot \xi} 
\lom W_1(\infty,\xi) \psi_i(\xi) \hat{\vect{P}}^\mu\, \overline{\psi}_j(0) W_2(0,\infty) \rom \\
\nonumber
& = \int \frac{d^4 \xi}{(2\pi)^4} e^{\i k \cdot \xi} \,
\i \frac{\partial}{\partial \xi_\mu}
\bigg\{ \lom W_1(\infty,\xi) \psi_i(\xi) \overline{\psi}_j(0) W_2(0,\infty) \rom \bigg\} \ ,
\end{align}
where, for brevity, we have omitted the color traces. The last step can be justified as follows:
\begin{align}
\label{e:comm_prop_P}
\lom W(\infty,\xi) \psi(\xi) \hat{\vect{P}}^\mu & = 
\lom \big[ W(\infty,\xi) \psi(\xi)\, , \hat{\vect{P}}^\mu \big] = 
\lom W(\infty,\xi) \big[ \psi(\xi)\, , \hat{\vect{P}}^\mu \big] + 
\lom \big[ W(\infty,\xi)\, , \hat{\vect{P}}^\mu \big] \psi(\xi) \\
\nonumber
& = \lom W(\infty,\xi) \bigg( \i \frac{\partial}{\partial \xi_\mu} \psi(\xi) \bigg) + 
\lom \bigg( \i \frac{\partial}{\partial \xi_\mu} W(\infty,\xi) \bigg) \psi(\xi) = 
\i \frac{\partial}{\partial \xi_\mu} \bigg\{ \lom W(\infty,\xi) \psi(\xi) \bigg\} \ . 
\end{align}
Finally,  
integrating by parts, we obtain the master result of this section,
\begin{equation}
\label{e:master_sum_rule}
\sum_h \sum_{S} 
\int \frac{d^4 P}{(2\pi)^4}\, (2\pi) \delta(P^2 - M_h^2) 
P^\mu \Delta^h(k,P,S) = 
k^\mu\, \Xi^{n.c.}(k) \ , 
\end{equation}
where the boundary terms have vanished because of the boundary conditions for the fermionic fields, and the {\it n.c. (no cut)} label at the r.h.s. means that we are not calculating the discontinuity of the inclusive jet correlator. 
It is important to notice that the derivation of Eq.~\eqref{e:master_sum_rule} holds true even with the (anti)time-ordering operators explicit, namely the specified choice for $W_{1,2}$ is not a necessary condition for the sum rule (but it is necessary for the spectral representation of $\Xi$ discussed in Section~\ref{ss:spectr_dec}).

The master sum rule~\eqref{e:master_sum_rule} encodes the connection between the quark-to-single-hadron fragmentation correlator and the jet correlator without the discontinuity, which coincides with the gauge invariant color averaged dressed quark propagator. The Dirac projections of its discontinuity give rise to the sum rules for collinear and TMD fragmentation functions that will be discussed in detail in the remainder of this section.
Note that, since Eq.~\eqref{e:master_sum_rule} involves a sum over the hadron spins, all the polarized structures in the $\Delta^h$ correlator vanish. Thus, we will be able to prove sum rules for unpolarized fragmentation functions only. 
Preliminary results on these FF sum rules have been presented at various conferences~\cite{Accardi:2018gmh}, and the unpolarized case has been discussed in Ref.~\cite{Accardi:2019luo}.

\subsection{Sum rules for rank 0 fragmentation functions}
\label{ss:mu_minus}

We now specialize the master sum rule~\eqref{e:master_sum_rule} to the TMD case in the parton frame. We start with the rank 0 term, defined in Eq.~\eqref{e:1hDelta_r0_r1}, which can be selected by choosing $\mu=-$.
We then consider the discontinuity of the sum rule, integrate both sides on the suppressed plus component of the partonic momentum, and choose the parton frame ($k_T = 0$).
We also exploit the relation
\begin{equation}
\label{e:measure}
\int \frac{d^4 P}{(2\pi)^3} \delta(P^2 - M_h^2) = 
\int \frac{dP^- d^2 \vect{P}_T}{2P^- (2\pi)^3} = 
\int \frac{dz d^2 \vect{P}_T}{2z (2\pi)^3} \ ,
\end{equation}
and obtain: 
\begin{equation}
\label{e:long_sr_operator_1}
\sum_{h\, S} \int \frac{dz d^2 \vect{P}_T}{2z (2\pi)^3} P^- \int dk^+ \,
\text{Disc}\, [\Delta^h(k,P,S)]_{\begin{subarray}{l} P^- = z k^- \\ k_T = 0 \end{subarray}} = 
k^- \int dk^+\, \text{Disc}\, [\Xi^{n.c.}(k)]_{k_T = 0} \ .
\end{equation}
This equation can be rewritten in terms of the collinear fragmentation correlator~\eqref{e:1hDelta_integrated} and the jet correlator Eq.~\eqref{e:J_TMDcorr}.  
The result is:
\begin{align}
\label{e:long_sr_operator_2}
\sum_{h\, S} \int dz\, z\, \Delta^h(z) = 
\sum_{h\, S} \int dz\, d^2 \vect{P}_T \, z\, \Delta_0^h(z,P_T) = 
2(2\pi)^3 J(k^-,\vect{0}_T) \ .
\end{align}
Note that only the hadron spin-independent part of $\Delta^h(z)$ survives in~\eqref{e:long_sr_operator_2}. 
Considering now the Dirac projections of the correlators on both sides, we can turn this into momentum sum rules for the collinear FFs in Eq.~\eqref{e:1hDelta_coll}. 
There are, in general, 9 Dirac projections $\Delta^{[\Gamma]}$ and $J^{[\Gamma]}$ 
involving twist-2 and twist-3 functions, of which only three are relevant for the rank 0 case:\footnote{The structures $\Gamma=\{ \gamma^-\gamma_5 ,\, \i\gamma_5 ,\,  \i\sigma^{-+}\gamma_5 \}$ project polarized TMD fragmentation functions out of $\Delta$. Since we are summing over the hadron polarization states in Eq.~\eqref{e:average_Ph}, these contributions vanish; on the contrary, these structure do not appear in $J$ from the very beginning because of parity invariance. The projections for the other 3 Dirac structures $\Gamma=\{ \i\sigma^{i-}\gamma_5 ,\, \gamma^i ,\, \gamma^i\gamma_5 \}$ produce the trivial result $0=0$.} 
\begin{align}
\label{e:sumrule_D1}
[\, \Gamma = \slashed{n}_+ \, ] \ \ \ \ 
& \sum_{h\, S} \int dz  z\,  D_1^{h}(z) = 1 \ ,	\\
\label{e:sumrule_E}
[\, \Gamma = \id \, ] \ \ \ \
& \sum_{h\, S} \int dz M_h E^{h}(z) = M_j \ , \\
\label{e:sumrule_H}
[\, \Gamma = \i \sigma^{\mu\nu} {n_i}_\mu {n_j}_\nu \gamma_5  \, ] \ \ \ \ 
& \sum_{h\, S} \int dz  M_h H^{h}(z) = 0 \ .
\end{align} 
To obtain the result for the collinear $D_1$ and $E$ FFs we have used Eq.~\eqref{e:alpha_calc_2} and~\eqref{e:zeta_calc_2} with $\theta(k^-)=1$ because $k^-$ is positive by four momentum conservation ($k^-$ is equal to the sum of the minus momenta of all produced hadrons, and these are physical on-shell particles). 

Renormalization is known to preserve Eq.~\eqref{e:sumrule_D1}~\cite{Collins:1981uw,Collins:2011zzd}.
The renormalization of $E^h(z)$ and its moments has been discussed in Refs.~\cite{Belitsky:1996hg,Belitsky:1997ay} at leading order in the strong coupling and in the large $N_c$ limit, using the light-cone gauge and neglecting current quark mass contributions. 
One can then infer an approximate evolution equation for $M_j$.
Instead, the renormalization of $H^h$, which is directly connected to a three-parton correlation function~\cite{Metz:2016swz}, has not yet been addressed to our knowledge.
Nevertheless, the derivations of these sum rules are rooted in the conservation of the partonic four-momentum encoded in Eq.~\eqref{e:master_sum_rule} and in the symmetry properties of the correlators $\Xi$ and $\Delta^h$. Therefore, we expect Eqs.\eqref{e:sumrule_D1}-\eqref{e:sumrule_H} and all the other sum rules discussed in this paper to be valid \emph{in form} also at the renormalized level in perturbative QCD.

The normalization~\eqref{e:rho13_positivity_rho3sumrule} of the spectral function $\rho_3$ -- which is a direct consequence of the equal-time anticommutation relations for the fermion fields~\cite{Weinberg:1995mt,Solis:2019fzm} -- is crucial to obtain the well-known {\it momentum sum rule}~\eqref{e:sumrule_D1} for the unpolarized fragmentation function $D_1^h$, that was originally proven without reference to the jet correlator~\cite{Collins:1981uw,Mulders:1995dh}. 
An experimental verification of this sum rule is therefore also an indirect check of the validity of the K\"allen-Lehman spectral representation. 
%
%
It is also interesting to note that Eq.~\eqref{e:sumrule_D1} allows one to write the unpolarized ``inclusive jet function'' and ``energy-energy correlation jet function'' introduced in the context of Soft Collinear Effective Theory as, respectively, the matching coefficients of the fragmenting jet functions onto the collinear FFs~\cite{Procura:2009vm,Jain:2011xz} and of the TMD FFs onto the collinear FFs~\cite{Moult:2018jzp,Luo:2019hmp}.


The chiral-odd sum rule~\eqref{e:sumrule_E}, that generalizes the one discussed in Ref.~\cite{Jaffe:1993xb,Jaffe:1996zw}, can be called {\em ``mass sum rule''} because of its physical interpretation: the non-perturbative jet mass $M_j$ corresponds to the sum of the masses of all possible particles produced in the hadronization of the quark, weighted by the chiral-odd collinear twist-3 fragmentation function $E^h(z)$. 
In looser terms, $M_j$ can be interpreted as the average mass of the hadronization products.

Finally, the sum rule~\eqref{e:sumrule_H} is, to our knowledge, new.

\subsection{Sum rules for rank 1 fragmentation functions}
\label{ss:mu_transverse}

Let us now specify the master sum rule to the case of the rank 1 correlator $\Delta_{1}^{h\, \alpha}(z,P_T)$ defined in Eq.~\eqref{e:1hDelta_r1}. This can be selected by choosing  $\mu = \alpha = 1,2$ in Eq.~\eqref{e:master_sum_rule}. Since we are working in the parton frame  
where $k_T=0$, this reads:
\begin{equation}
\label{e:sumrule_transverse}
\sum_{h\, {S}} \int dz\ d^2 \vect{P}_T\, P_T^\alpha\, \Delta^h(z,P_T) = 
\sum_{h\, {S}} \int dz\ d^2 \vect{P}_T\, P_T^\alpha\, {P_T}_\rho\, \Delta_1^{h\, \rho}(z,P_T) =
0 \, . 
\end{equation}
This result can also be achieved directly from Eq.~\eqref{e:unintegrated_sum_rule} choosing the parton frame, performing the integration explicitly with $\mu$ transverse index, and assuming that the fermion fields vanish at the boundary of space~\cite{Meissner:2010cc}. 

Using the correspondence between symmetric traceless tensors built with the transverse momentum and complex numbers outlined in Appendix~\ref{a:conv} and Ref.~\cite{Boer:2016xqr}, and the relation~\cite{Bacchetta:2006tn,Mulders:2016pln}
\begin{equation}
\label{e:NT_eps_relation}
P_T^{[ \alpha} \epsilon_T^{\rho\sigma ]} {P_T}_\rho = P_T^2 \epsilon_T^{\alpha\rho} \ ,
\end{equation}
we can calculate the Dirac projections of Eq.~\eqref{e:sumrule_transverse}, based on the parametrization given in Eq.~\eqref{e:1hDelta_r1}. The result reads
\begin{align}
\label{e:sumrule_H1p}
[\, \Gamma = -\i \sigma^{\mu\nu} {n_i}_\mu {n_+}_\nu \gamma_5 \, ] \ \ \ \ 
& \sum_{h\, S} \int dz z M_h\, H_1^{\perp\, (1)\, h}(z) = 0 \ , \\
\label{e:sumrule_Dp}
[\, \Gamma = -\slashed{n}_i \, ] \ \ \ \ 
& \sum_{h\, S} \int dz M_h^2\, D^{\perp\, (1)\, h}(z) = 0 \ , \\
\label{e:sumrule_Gp}
[\, \Gamma = -\slashed{n}_i \gamma_5 \, ] \ \ \ \ 
& \sum_{h\, S} \int dz M_h^2\, G^{\perp\, (1)\, h}(z) = 0 \ , 
\end{align}
where we defined the first $P_T$-moment of a generic fragmentation function $D$ as (see Appendix~\ref{a:frame_dep}): 
\begin{equation}
\label{e:def_Php_mom}
D^{(1)}(z) = \int d^2 \vect{P}_T \frac{\vect{P}_T^2}{2 z^2 M_h^2} D(z,P_T^2) \ .
\end{equation}
As in Section~\ref{ss:mu_minus}, the remaining Dirac projections yield the trivial result $0=0$. The sum rule~\eqref{e:sumrule_H1p} for $H_1^{\perp\, h}$ is also known as the Sch\"afer-Teryaev sum rule~\cite{Schafer:1999kn,Meissner:2010cc}. 
We already discussed the sum rule for $D^{\perp\, h}$ in Ref.~\cite{Accardi:2019luo}, and that for $G^{\perp\, h}$ is new.
The QCD evolution of the first moment of $H_1^{\perp\, h}$ has been discussed in Ref.~\cite{Kang:2010xv}, but no statement is available on the validity of the Sch\"afer-Teryaev sum rule under renormalization. Nevertheless, the sum rule for the T-odd FFs, $H_1^{\perp\, h}$, $G^{\perp\, h}$, and $H^h$, is a result of the absence of T-odd terms in the inclusive jet correlator, a feature that should be preserved under renormalization, too. Checking this by explicit calculation remains an interesting exercise for the future.


\subsection{Sum rules for dynamical twist-3 fragmentation functions}
\label{ss:qgq_sumrules}

Let us now consider the equations of motion relations (EOMs) which relate twist-2 and twist-3 fragmentation functions in the parton frame:
\begin{align}
  \label{e:eom_E}
  & E^h = \widetilde{E}^h + z \frac{m}{M_h} D^h_1 \\
  \label{e:eom_H}
  & H^h = \widetilde{H}^h - \frac{\vect{P}_T^2}{z M_h^2} H_1^{\perp\, h} \\
  \label{e:eom_Dp}
  & D^{\perp\, h} = \widetilde{D}^{\perp\, h} + z D^h_1 \\
  \label{e:eom_Gp}
  & G^{\perp\, h} = \widetilde{G}^{\perp\, h} + z \frac{m}{M_h} H_1^{\perp\, h} \ ,
\end{align}
where the functions with a tilde parametrize the twist-3 $\widetilde{\Delta}_A^\alpha$ quark-gluon-quark correlator~\cite{Bacchetta:2006tn}, and $m$ is the current mass of the specific quark considered. These relations, that are a consequence of the Dirac equation for the quark field, have been originally presented in the hadron frame~\cite{Tangerman:1994bb,Mulders:1995dh} (see also Ref.~\cite{Bacchetta:2006tn}) and in Appendix~\ref{a:frame_dep} we discuss their transformation to the parton frame. 

The Eqs.~\eqref{e:eom_E}-\eqref{e:eom_Gp} allow us to investigate the momentum sum rules for the ``dynamical" twist-3 FFs (those with a tilde) without explicitly working with  the quark-gluon-quark fragmentation correlator $\widetilde{\Delta}_A^\alpha$ and the quark-gluon-quark inclusive jet correlator $\widetilde{J}_A^\alpha$ introduced in Ref.~\cite{Accardi:2017pmi}.
Indeed, combining the four EOMs with the sum rules discussed in Section~\ref{ss:mu_minus} and Section~\ref{ss:mu_transverse} we obtain
\begin{align}
\label{e:sumrule_Et}
& \sum_{h\, S} \int dz M_h \widetilde{E}^{h}(z) = M_j - m = m^{\text{corr}}\\
\label{e:sumrule_Ht}
& \sum_{h\, S} \int dz M_h \widetilde{H}^{h}(z) = 0 \\
\label{e:sumrule_Dtp}
& \sum_{h\, S} \int dz M_h^2 \widetilde{D}^{\perp\, (1)\, h}(z) = 
- \sum_{h\, S} \int dz z\, M_h^2 D_1^{(1)\, h}(z) \ \equiv\
\frac{1}{2} \langle P_{T}^2 / z^2 \rangle \\
\label{e:sumrule_Gtp}
& \sum_{h\, S} \int dz M_h^2 \widetilde{G}^{\perp\, (1)\, h}(z) = 0 \ ,
\end{align}
which provide a complete set of sum rules for the four unpolarized dynamical twist-3 FFs. As with the twist-2 case, no sum rule can be established for polarized FFs.

Eq.~\eqref{e:sumrule_Et} is the generalization of the sum rule $\int dz \widetilde{E} = 0$ discussed in Ref.~\cite{Bacchetta:2006tn}. 
This generalization is based on the fact that the jet mass $M_j$ differs in general from the current quark mass by an amount $m^{\text{corr}}$ (the correlation mass introduced in Eq.~\eqref{e:Mj_decomp}) which we argued is non-perturbatively generated by quark-gluon-quark correlations and the dynamical breaking of the chiral symmetry. In the FF correlator $\widetilde{\Delta}_A^\alpha$, the chiral odd component of these correlations is parametrized by the $\widetilde E$ function~\cite{Bacchetta:2006tn}, that also provides a flavor decomposition for $m^{\text{corr}}$ through Eq.~\eqref{e:flav_mqcorr}. See Section~\ref{ss:sumrules_summary} for a deeper discussion of this point. 

The sum rule~\eqref{e:sumrule_Dtp} connects the first moment of the twist-3 $\widetilde D^\perp$ FF to the average squared transverse momentum acquired by unpolarized hadrons fragmented from an unpolarized quark~\cite{Accardi:2019luo} (for the definition of the average operator see Appendix~\ref{a:frame_dep}). Therefore, this sum rule also probes the nature of the non-perturbative hadronization process in analogy with the way the sum rule for $\widetilde E$ probes the nature of the vacuum. 
Similar relations exist in literature, see e.g. Eq.~(76) in Ref.~\cite{Metz:2016swz}, which connects a three-parton FF, the first moment of the Collins FF, and the average transverse momentum of an unpolarized hadron fragmenting from a transversely polarized quark. Another example is the relation between the average transverse momentum of an unpolarized quark in a transversely polarized hadron and the Qiu-Sterman function~\cite{Qiu:1991pp,Qiu:2020oqr}. 

It is finally worthwhile remarking that the sum rules for $D^\perp$ and $\widetilde{D}^\perp$ are frame dependent because such are the involved transverse momenta. On the contrary, all other sum rules are frame independent. Appendix~\ref{a:frame_dep} discusses these features in details.

\section{Sum rules compendium and discussion}
\label{ss:sumrules_summary}

We collect here for convenience the complete set of sum rules for twist-2 and twist-3 FFs scattered throughout Section~\ref{s:1h_rules}, and remind that all fragmentation functions implicitly depend on the quark flavor, omitted for sake of simplicity.
%
%
At twist 2,
\begin{align}
\label{e:pframe_sumrule_D1}
& \sum_{h\, S} \int dz\, z\, D_1^{h}(z) = 1 \, , \\
\label{e:pframe_sumrule_H1p}
& \sum_{h\, S} \int dz\, z\, M_h H_1^{\perp\, (1)\, h}(z) = 0 \, .
\end{align}
%
%
At twist 3,
\gdef\thesubequation{\theequation \textit{a,b}}
\begin{subeqnarray}
\label{e:pframe_sumrule_E_Et}
& \ \ \displaystyle \sum_{h\, S} \int dz\, M_h\, E^{h}(z) = M_j\, , \qquad \ \
& \sum_{h\, S} \int dz\, M_h\, \widetilde{E}^{h}(z) = M_j - m = m^{\text{corr}}\, , \\
\refstepcounter{equation}
\label{e:pframe_sumrule_H_Ht}
& \displaystyle \sum_{h\, S} \int dz\, M_h\, H^{h}(z) = 0\, , \qquad \ \ 
& \sum_{h\, S} \int dz\, M_h\, \widetilde{H}^{h}(z) = 0\, , \\
\refstepcounter{equation}
\label{e:pframe_sumrule_Dp_Dpt}
& \ \ \ \ \ \ \displaystyle\sum_{h\, S} \int dz\, M_h^2\,  D^{\perp\, (1)\, h}(z) = 0\, , \qquad \ \  
& \sum_{h\, S} \int dz\, M_h^2\, \widetilde{D}^{\perp\, (1)\, h}(z) = \frac{1}{2} \langle P_{T}^2 / z^2 \rangle\, , \\
\refstepcounter{equation}
\label{e:pframe_sumrule_Gp_Gpt}
& \ \ \ \ \ \ \displaystyle\sum_{h\, S} \int dz\, M_h^2\, G^{\perp\, (1)\, h}(z) = 0\, , \qquad \ \  
& \sum_{h\, S} \int dz\, M_h^2\, \widetilde{G}^{\perp\, (1)\, h}(z) = 0 \, .
\end{subeqnarray}

The sum rules~\eqref{e:pframe_sumrule_D1},~\eqref{e:pframe_sumrule_H1p} and~\eqref{e:pframe_sumrule_H_Ht}$b$ were already known in literature~\cite{Collins:1981uw,Mulders:1995dh,Meissner:2010cc,Schafer:1999kn,Bacchetta:2006tn}, with the latter proven here for the first time at the correlator level. 
It's interesting to notice that the sum rules for the T-odd FFs $H^h$, $H_1^{\perp\, h}$, $G^{\perp\, h}$ are a consequence of the absence of T-odd terms in the inclusive jet correlator~\eqref{e:invariant_quark_correlator} (see Eq.~\eqref{e:B3_zero}). Being the consequence of time-reversal symmetry, this feature is frame-independent (see Appendix~\ref{a:frame_dep}).
The sum rules~\eqref{e:pframe_sumrule_E_Et} for $E$ and $\widetilde E$ have been originally discussed in Ref.~\cite{Jaffe:1996zw}, but here extended to the non-perturbative domain (we will have more to say about these shortly). 
All others are, to the best of our knowledge, novel results\footnote{A partial proof was discussed in Ref.~\cite{Accardi:2019luo} and at various conferences, see for example Ref.~\cite{Accardi:2018gmh}.}. \\ 

%

As we will discuss below, these sum rules are generically useful as constraints in phenomenological fits where experimental data is scarce, and when developing fragmentation models.
The non-zero sum rules, however, have a significance that goes well beyond that. 
To start with, we have shown that the $D_1$ sum rule is theoretically linked to the normalization property~\eqref{e:rho13_positivity_rho3sumrule} of the chiral-even $\rho_3$ spectral function, and, thus, to the equal-time anticommutation relations for the fermion fields. Hence its experimental verification also entails an indirect check of the validity of the K\"allen-Lehman spectral representation. 
The sum rules~\eqref{e:pframe_sumrule_E_Et} for $E$ and $\widetilde E$, and~\eqref{e:pframe_sumrule_Dp_Dpt}$b$ for $\widetilde D^\perp$ are also noteworthy because, unlike the others, they are sensitive to aspects of the non perturbative dynamics of QCD: respectively, the dynamical mass generation in the QCD vacuum, and the transverse momentum generation in the fragmentation process~\cite{Accardi:2019luo}. 

Our proof has been developed in the parton frame in order to connect to the inclusive jet correlator, that cannot be defined in the hadron frame.
Most of the sum rules are nonetheless frame independent, as detailed in Appendix~\ref{a:frame_dep}. The only exceptions are the sum rules~\eqref{e:pframe_sumrule_Dp_Dpt} for $D^\perp$ and $\widetilde{D}^\perp$, that in the hadron frame exchange the role of the kinematic and dynamical twist-3 functions. In that frame, it is the first moment of the $D^{\perp h}$ functions that are sensitive to the transverse momentum of the fragmented hadrons, whereas the first moment of the $\widetilde{D}^{\perp h}$ functions sum up to zero. 

Finally, note that our proofs are at present valid only for unrenormalized FFs. However, the arguments we utilized are rooted in the conservation of the partonic four-momentum encoded in Eq.~\eqref{e:master_sum_rule}, and on the symmetry properties of the correlators $\Xi$ and $\Delta^h$. For this reason, we expect all momentum sum rules to be valid in form also at the renormalized level. 
In fact: renormalization is known to preserve Eq.~\eqref{e:pframe_sumrule_D1}~\cite{Collins:2011zzd,Collins:1981uw}; 
the evolution of $E(z)$ and $M_j$ can be inferred from the results presented in Refs.~\cite{Belitsky:1996hg,Belitsky:1997ay}; it could also be argued that the sum rules for the T-odd FFs are preserved under renormalization due to the absence of T-odd terms in the inclusive jet correlator, but explicit calculations are needed to corroborate this hypothesis and to understand the behavior of all the other sum rules.  


\subsection{Dynamical chiral symmetry breaking}
\label{sss:DCSB}

The mass sum rules~\eqref{e:pframe_sumrule_E_Et} are of particular interest, because they shed additional light on the QCD mass generation mechanism already explored in Section~\ref{s:jetcor} in terms of the jet correlator $J$ and its chiral odd component. As discussed in Section~\ref{ss:TMDjet_recap}, the jet mass $M_j=m+m^{\text{corr}}$ quantifies the dressing of a quark as it propagates in the QCD vacuum. Here we suggest that, whereas the current quark mass $m$ is the component of $M_j$ that explicitly breaks the chiral symmetry, it is the  $m^{\text{corr}}$ correlation mass component that can be considered a theoretically solid order parameter for its dynamically breaking.  That this is the case is supported by the following arguments, highlighting the central role played by quark-gluon interactions in generating $m^{corr}$, and how this is intrinsically connected to the properties of the QCD vacuum. 

Overall, the correlation mass can vanish in two circumstances, where the neglect of quark-gluon-quark correlations is achieved in different ways. In the first case, one can invoke the ``Wandzura-Wilczek (WW) approximation'', which consists in neglecting the twist-3 ``tilde'' functions, that parametrize the strength of quark-gluon-quark correlations, compared to the twist-2 and twist-3 functions without a tilde, that describe quark-quark correlations. In other words, this approximation consists in neglecting the role of gluons except in the dressing of the quark-quark correlators, and setting the ``tilde'' functions to zero~\cite{Bastami:2018xqd}\footnote{The WW appriximation takes its name from the fact that one is in fact utilizing simplified form of the Wandzura-Wilczek-type relations, originally introduced and discussed in Ref.~\cite{Wandzura:1977qf}, that relate twist-2 and twist-3 functions. While in some processes the neglect of quark-gluon-quark interactions leads to phenomenologically successful comparisons to experimental data \cite{Bastami:2018xqd}, this assumption is not a priori justified in all circumstances~\cite{Accardi:2017pmi}. In particular, one needs to make sure that the dominant quark-quark terms do not cancel in the observable of interest. For example, in our case, a WW approximation applied to Eq.~\eqref{e:sumrule_Dtp} would amount to predicting no transverse momentum in the fragmentation process, $\langle P_{T}^2 / z^2 \rangle = 0$, which is clearly not the case.}
Thus, $M_j \overset{\,_{WW}}{=} m$ and $m^{\text{corr}} \overset{\,_{WW}}{=} 0$, as can be easily seen by setting $\widetilde E = 0$ in Eq.~\eqref{e:eom_E} and using the sum rules~\eqref{e:sumrule_D1} and~\eqref{e:sumrule_E}. (An application of the same WW approximation to the sum rule~\eqref{e:sumrule_Et} consistently provides one with the identity $0=0$.)
Another case in which the dynamical mass $m^{\text{corr}}$ vanishes is when the non-interacting vacuum $|0 \rangle$ of the theory is used in place of the interacting one $\rom$~\cite{Peskin:1995ev}, so that one cannot fully contract the $\psi A_T^\alpha \bar\psi$ operator that defines $\widetilde E$ unless the interaction terms in the Lagrangian are taken into account, effectively causing $\widetilde E = 0$ as in the WW approximation. 

Furthermore, one can decompose the chiral-odd spectral function $\rho_1$ into a pole part, with an isolated singularity at the (renormalized) current mass value $\mu^2=m^2$, and a remnant $\overline\rho_1$ (see, {\it e.g.}, Ref.~\cite{Solis:2019fzm}):
\begin{align}
  \rho_1(\mu^2) = \delta(\mu^2-m^2) + \overline\rho_1(\mu^2) \ .
\end{align}
This singularity, in fact cannot appear in the full, non-perturbative propagator because quarks are not physical states of the theory; rather, it originates from a perturbative treatment of a the propagating quark considered as an asymptotic field while in reality it is not. 
Next, combining the $E$ and $\widetilde E$ sum rule with the EOM relation~\eqref{e:eom_E} one sees that
\begin{align}
\label{e:nonpole_rho1_mcorr}
	m^{\text{corr}} \overset{lcg}{=} \int d\mu^2 \sqrt{\mu^2}  \, \overline\rho_1(\mu^2) \ .
\end{align}
Therefore, the perturbative pole 
is effectively removed in the sum rule for $\widetilde E$, and from the spectral decomposition of the correlation mass. This is all the more interesting, because it is the twist-3 $\widetilde E$ function rather than $E$, that contributes to hadroproduction DIS processes, whence the sum rules can be experimentally measured~\cite{Bacchetta:2006tn}.

Finally,  the $\widetilde E$ sum rule also provides one with a hadronic flavor decomposition of the correlation mass: 
\begin{equation}
\label{e:flav_mqcorr}
m^{\text{corr}} = \sum_{h,S} \int dz M_h\, \widetilde{E}^{h}(z) 
  \equiv \sum_h \, m_h^{\text{corr}} \ ,
\end{equation}
where each $m_h^{\text{corr}} = \sum_{S} \int dz M_h\, \widetilde{E}^{h}(z)$ quantifies the contribution to the interaction-dependent part of the jet mass associated to the hadronization into a specific hadron $h$. One can therefore envisage investigating the separate role of baryon and light mesons in the dynamical chiral symmetry breaking process, with the pions and kaons expected to become massless in the chiral limit due to the Goldstone theorem, and obtain a more fine-grained picture of the spontaneous generation of mass in QCD.
As a starter, calculations of $E$ and $\widetilde{E}$ in models which incorporate the dynamical breaking of the chiral symmetry, for example such as in treatments combining the Nambu--Jona-Lasinio model~\cite{Ito:2009zc,Matevosyan:2011vj,Bentz:2016rav} 
The full set of sum rules provided in this article could be used to constrain and refine these calculations, and a comparison with even a limited amount of experimental data on $\widetilde E$ would provide the model with the dynamical input necessary to explore with confidence the chiral limit via Eq.~\eqref{e:flav_mqcorr}.

\subsection{Phenomenology}
\label{sss:pheno}

These sum rules can be of phenomenological relevance in the studies of hard scattering process with hadrons in the final states, 
for example, semi-inclusive deep-inelastic scattering (SIDIS) and electron-positron annihilation into one or two hadrons, 
as well as hadroproduction in hadronic collisions at both fixed target and collider facilities~\cite{Hadjidakis:2018ifr,Aidala:2019pit}).

The leading-twist TMD FFs $D_1$ and $H_1^\perp$ can be observed in 
SIDIS procceses, considering specific angular modulations of the cross section at low transverse momentum~\cite{Bacchetta:2006tn}. 

The dynamical twist-3 FFs ($\widetilde E$, $\widetilde H$, $\widetilde D^\perp$, $\widetilde G^\perp$) appear in the SIDIS cross section at order $1/Q$, where $Q$ is the hard scale of the process. As it turns out, these are the only twist-3 FFs contributing to the cross section in a frame where the azimuthal angles refer to the axis given by the four-momenta of the target nucleon and the photon, rather than of the target nucleon and the detected hadron~\cite{Bacchetta:2006tn}.
In such a frame, their kinematic twist-3 counterparts (those without a tilde in their symbol) do not contribute to the  cross section, but can be obtained from the former by use of the equation of motion relations \eqref{e:eom_E}-\eqref{e:eom_Gp}. 
The role of twist-3 FFs in other semi-inclusive processes is reviewed in Ref.~\cite{Metz:2016swz}. 
In general, to access these fragmentation functions one needs to calculate cross sections at least to twist-3 level, and, in the case of the chiral-odd $E$ and $\widetilde E$ FFs, to combine these with another chiral-odd distribution or fragmentation function.

The possibility to experimentally observe the tilde functions in semi-inclusive processes is particularly interesting for the case of $\widetilde E$, which contributes to the determination of the interaction-dependent correlation mass $m^{\text{corr}}$ and its flavor decomposition through the sum rule \eqref{e:pframe_sumrule_E_Et}. This is not, however, the only experimental window on $m^{\text{corr}}$. For example, as discussed in Ref.~\cite{Accardi:2017pmi,Accardi:2018gmh}, the correlation mass $m^{\text{corr}}$ also contributes coupled to the collinear transversity PDFs to the inclusive DIS $g_2$ structure function at large Bjorken $x_B$. Likewise, the correlation mass couples to the collinear transversity FF $H_1$ in single hadron production of, say, the self-polarizing $\Lambda$ particle in semi-inclusive $e^+e^-$ collisions. Likewise, it can couple to the dihadron $H_1^\sphericalangle$ FF in the case of same-hemisphere double hadron production. 

As one can see, the experimental information we are after is scattered among a umber of diverse observables and process. One way to gather it in a consistent fashion is to perform ``universal'' QCD fits of a suitable subsets of PDFs and FFs. One possibility is to simultaneously fit $m^{\text{corr}}$, the collinear transversity PDFs $h_1$, and the collinear dihadron $H_1^\sphericalangle$. The needed processes are longitudinal-transverse asymmetries in inclusive DIS ($\propto m^{\text{corr}}\, h_1 $~\cite{Accardi:2017pmi}), 
di-hadron production in SIDIS ($\propto h_1 H_1^\sphericalangle$~\cite{Radici:2015mwa,Radici:2018iag}), the Artru-Collins asymmetry in double di-hadron production in electron-positron annihilation ($\propto H_1^\sphericalangle H_1^\sphericalangle$~\cite{Matevosyan:2018icf}), and semi-inclusive same-side dihadron production ($\propto m^\text{corr} H_1^\sphericalangle$~\cite{Accardi:2017pmi}). This kind of universal QCD analysis, seeking to numerically fit several non perturbative functions at once, is numerically very demanding in terms of raw computational power and stability of the fitting algorithms. Nonetheless its feasibility has been recently demonstrated in a series of works by the JAM collaboration~\cite{Ethier:2017zbq,Lin:2017stx,Sato:2019yez}. 

In order to properly separate perturbative and non-perturbative contributions, these observables should be addressed in the context of the associated factorization theorems. In this respect, resummed perturbative QCD and Soft-Collinear Effective Theories (SCET) provide the needed tools. Namely, the inclusive jet correlator $\Xi$ emerges, {\em e.g.}, in the factorization of the so-called end-point region of DIS processes at large $x$~\cite{Becher:2006mr,Becher:2006nr,Chen:2006vd,Sterman:1986aj,Chay:2005rz}, where the final state invariant mass $Q(1-x) \sim \Lambda_{\text{QCD}}$, and $Q$ is the hard momentum transfer. Those analyses should be extended to the chiral-odd components of the jet correlator, and also applied to SIDIS and $e^+e^-$ annihilation into one or two hadrons.

\section{Summary and outlook}
\label{s:conclusions}


In this paper we have studied the properties of the fully inclusive jet correlator~\eqref{e:invariant_quark_correlator} introduced and in, {\em e.g.}, Ref.~\cite{Sterman:1986aj,Chen:2006vd,Collins:2007ph,Accardi:2008ne,Accardi:2017pmi,Accardi:2019luo}. 
In particular, in Section~\ref{sss:link_structure} we have have presented a gauge-invariant definition for this correlator, and discussed a specific class of Wilson lines (staple-like) that allows one to re-write this as the gauge-invariant quark propagator~\eqref{e:invariant_quark_correlator_W}.
Moreover, in Section~\ref{sss:dirac_structure} we have decomposed the fully inclusive jet correlator in Dirac structures, and organized the various terms according to their suppression in powers of $\Lambda/k^-$, where $\Lambda$ is a generic hadronic scale and $k^-$ the dominant light-cone component of the quark momentum. 

As a byproduct of the Dirac decomposition of the jet correlator, we have provided a gauge invariant definition for the inclusive jet mass $M_j$, an object which encodes the physics of the hadronizing color-averaged dressed quark. 
This mass can be decomposed in terms of the current quark mass and a dynamical component generated by nonperturbative quark-gluon-quark correlations (see Eq.~\eqref{e:Mj_decomp}). 
New non-perturbative effects induced by this mass and its dynamical component can emerge at the twist-3 level, for example in inclusive deep-inelastic scattering at the level of the $g_2$ structure function~\cite{Accardi:2017pmi,Accardi:2018gmh}, and potentially in semi-inclusive DIS, in semi-inclusive annihilation into one or two hadrons, and in hadronic collisions (see Section~\ref{sss:pheno}).

In Section~\ref{ss:spectr_dec}, we have developed a spectral representation for the gauge-invariant quark propagator, and we have connected the jet's mass and virtuality to the chiral-odd and even spectral functions, respectively. In particular, in the light-cone gauge the jet mass reduces to the first moment of the chiral odd spectral function, which provides a link to non-perturbative treatments of the quark propagator and in particular to the properties of the associated mass function~\cite{Siringo:2016jrc,Solis:2019fzm,Roberts:2007jh,Roberts:2015lja}, see Section~\ref{ss:TMDjet_recap}. In analogy with the role played by the dressed quark mass and the mass function, the dynamical component of the jet mass can be interpreted as an order parameter for the dynamical breaking of the chiral symmetry (see Section~\ref{ss:TMDjet_recap} and Section~\ref{sss:DCSB}). 

In Section~\ref{s:1h_rules}, we have presented a connection at the operator level between the single-hadron fragmentation correlator~\eqref{e:1hDelta_corr} and the fully inclusive jet correlator~\eqref{e:invariant_quark_correlator}. This connection, encoded in the master sum rule~\eqref{e:master_sum_rule}, provides an explicit link between the propagation of the quark and the fully inclusive limit of its hadronization. The chosen class of Wilson lines allows one to connect these operators to matrix elements accessible in high-energy scattering experiments.
In fact, from the master sum rule~\eqref{e:master_sum_rule} we have derived momentum sum rules for the fragmentation functions of quarks into unpolarized hadrons up to twist 3, confirming sum rules already known in the literature and proposing new ones (see Section~\ref{ss:sumrules_summary}). 
Among the others, the novel sum rules for the $\widetilde E$ and the $\widetilde{D}^\perp$ FFs have a dynamical interpretation: the RHS of this sum rules corresponds, respectively, to the mass and the average squared transverse momentum generated during the fully inclusive hadronization of a nearly massless quark.

Moreover, we have connected the sum rules for the $D_1$ and the $E$ FFs to the integral of the quark's chiral-even and chiral-odd spectral functions, whose integrals become experimentally measurable quantities (see Eq.~\eqref{e:sumrule_D1} and Eq.~\eqref{e:sumrule_E}, respectively). 
As a result, the sum rule for the unpolarized $D_1$ FF acquires a new deep interpretation, which goes beyond conservation of the collinear quark momentum: the RHS of the momentum sum rule for $D_1$ is precisely the normalization of the chiral-even $\rho_3$ spectral function, whose value only depends on the equal-time (anti)commutation relations for the fields involved~\cite{Weinberg:1995mt,Solis:2019fzm}. 
The mass sum rules for the $E$ and $\widetilde E$ FFs, instead, provide us with 
a way to constrain the chiral-odd $\rho_1$ spectral function, or, equivalently, to measure the color-screened dressed quark mass $M_j$ and its dynamical component $m^{\rm corr}$, respectively.
We believe that the possibility to experimentally access quantities connected to the dynamical breaking of the chiral symmetry in QCD is one of the most important outcomes of this paper.


\begin{acknowledgments}
We thank A. Bacchetta, I. Clo\"et, J. Goity, P.J. Mulders, J.W. Qiu, M. Radici, C. Roberts, C. Shi, A. Vladimirov 
for helpful discussions. 
This work was supported by the  U.S. Department of Energy contract DE-AC05-06OR23177, under which Jefferson Science Associates LLC manages and operates Jefferson Lab. 
AA also acknowledges support from DOE contract DE-SC0008791.
AS also acknowledges support from the U.S. Department of Energy, Office of Science, Office of Nuclear Physics, contract no. DE-AC02-06CH11357 and 
from the European Commission through the Marie Sk\l{}odowska-Curie Action SQuHadron (grant agreement ID: 795475). 
\end{acknowledgments}
\begin{appendix}

\section{Conventions}
\label{a:conv}

In this Appendix we discuss our light-cone conventions and notation. We start by reviewing the Fourier transform, the light-cone basis vectors for the longitudinal Minkowksi subspace, and the tensors needed to discuss parton and hadron dynamics in the transverse subspace. We then turn to  the conventions for the integration over the suppressed momentum component used to define the TMD fragmentation correlator and the fully inclusive jet correlators, and for the Dirac traces (or projections) needed to define the related TMD functions. For completeness, we also include a short discussion of the parton distribution correlator restricted to the spin-independent case.

\subsection{Fourier transform}

In order to be consistent with a large share of the literature dealing with TMD parton distribution and fragmentation functions 
(e.g. Refs.~\cite{Bacchetta:2006tn,Mulders:2016pln,Levelt:1993ac,Mulders:1995dh}), we define the Fourier transform with a $1/(2\pi)^4$ factor for space-time four vector integrations, see {\it e.g.} the definition of $\Xi(k;\omega)$ in Eq.~\eqref{e:invariant_quark_correlator}. Correspondingly, we do not include such factor in four-momentum integrations, contrary to, {\it e.g.}, Refs.~\cite{Bjorken:1965zz,Roberts:2015lja,Roberts:2007jh,Siringo:2016jrc,Solis:2019fzm}. This, in particular, results in an additional $1/(2\pi)^4$ factor in Eq.~\eqref{e:Feyn_spec_rep} and~\eqref{e:SF_mass} with respect to the definitions in Refs.~\cite{Bjorken:1965zz,Roberts:2015lja,Roberts:2007jh}.

\subsection{Light-cone coordinates and transverse space}

In a given reference frame, we collect the space-time components of a four-vector $a^\mu$ inside round parentheses, $a^\mu=(a^0,a^1,a^2,a^3)$, with $a^0$ the time coordinate. 
We define the light-cone $\pm$ components of the $a$ vector as
\begin{align}
  a^{\pm} = \frac{1}{\sqrt{2}} (a^0 \pm a^3) 
\end{align}
and collect these inside square brackets: $a^\mu = [a^-,a^+,\vect{a}_T]$, with $\vect{a}_T=(a^1,a^2)$ being the 2-dimensional components in transverse space. 
We also define the transverse four-vector as $a_T^\mu = [0,0,\vect{a}_T]$, such that $a_T^2 = -\vect{a}_T^2$. 
Namely, the norm of $\vect{a}_T$ is taken according to the Euclidean metric $\delta_T^{ij}=\text{diag}(1,1)$, whereas the norm of $a_T$ is calculated using the Minkowski metric $g^{\mu\nu}=\text{diag}(1,-1,-1,-1)$. 
Note that, in this paper, we consider highly boosted quarks and hadrons with dominant momentum component along the negative 3-axis, namely along the negative light-cone direction. Hence, we grouped the light-cone components inside the square parenthesis starting with the minus component.

The light-cone basis vectors are defined as:
\begin{equation}
\label{e:def_np_nm}
n_{\pm} = \frac{1}{\sqrt{2}} (1,0,0,\pm 1) \ \ , 
\end{equation}
such that $n_+^2 = n_-^2 =0$, $n_+^\mu n_{-\mu} = 1$, and $a^\pm = a^\mu {n_\mp}_\mu$.  Upon considering a specific process, the basis vectors $n_{\pm}^\mu$ can be determined by physical quantities. For example, in inclusive deep-inelastic scattering one can choose the four-momentum of the target and virtual photon to lie in the plus-minus plane; in semi-inclusive processes the tagged hadron's momentum can replace either one, typically the photon's momentum. In semi-inclusive electron-positron annihilation into two hadrons, one typically chooses the four-momenta of both tagged hadrons. In this paper, however, we consider quark propagation and fragmentation independently of any specific process, and will study the Lorentz transformation between the different frames in which the quark hadronization mechanism can be studied (see Appendix~\ref{a:frame_dep}). 

Following Ref.~\cite{Bacchetta:2006tn,Mulders:2016pln}, the transverse projector, $g_T^{\mu\nu}$, and the transverse anti-symmetric tensor, $\epsilon_T^{\mu\nu}$, are defined as:
\begin{align}
\label{e:gT_def}
g_T^{\mu\nu} & \equiv g^{\mu\nu} - n_+^{ \{ \mu} n_-^{\nu \} } \\ 
\label{e:epsT_def}
\epsilon_T^{\mu\nu} & \equiv \epsilon^{\mu \nu \rho \sigma} {n_-}_\rho {n_+}_\sigma  \equiv \epsilon^{\mu \nu + -} \ ,
\end{align}
where $g^{\mu\nu}$ is the Minkowski metric, $\epsilon^{\mu\nu\rho\sigma}$ is the totally anti-symmetric Levi-Civita tensor (with $\epsilon^{0123}=1$).
Note that $g_T^{\mu\nu} a_\nu = a_T^\mu$ projects a four-vector onto its transverse component, and $\epsilon_T^{\mu\nu} a_\nu = \epsilon_T^{\mu\nu} a_{T\nu}$ rotates that component by 90 degrees in the transverse plane.

In the paper we also make use of the correspondence between symmetric traceless tensors of definite rank and complex numbers, detailed in Ref.~\cite{Boer:2016xqr}. Its essence is the possibility to trade an uncontracted rank-$m$ tensor with a complex number. For a rank-$m$ tensor $T$ built out of a single transverse vector $\vect{a}_T$, this reads:
\begin{equation}
  \label{e:SST_C}
  T^{i_1 \cdots i_m}(\vect{a}_T) \rightarrow \frac{1}{2^{m-1}}\ 
  |\vect{a}_T|^m\ e^{\pm \i m \phi} \ ,
\end{equation}
where $\phi$ is the polar angle associated to $\vect{a}_T$ in the transverse plane. The rank $m$ of the tensor is reflected in the power of the modulus and in the phase of the complex number.

A useful consequence of this correspondence is that expressions proportional to $T^{i_1 \cdots i_m}$ vanish upon integration over $\vect{a}_T$, due to the angular part of the integration measure.
In our analysis we apply this correspondence to the following rank-2 tensor, built out of the hadron's transverse momentum $P_T$~\cite{Boer:2016xqr,vanDaal:2016glj}:
\begin{equation}
\label{e:STT2_Php}
P_T^{ij} \ \equiv\ P_T^i P_T^j + \frac{\vect{P}_T^2}{2} g_T^{ij} \ .
\end{equation}

\subsection{Parton distribution correlator}
The TMD parton distribution correlator and its Dirac projections are defined as:
\begin{equation*}
\label{e:proj_int_Phi}
\refstepcounter{equation}
\Phi(x,p_T)\ \equiv \int_{-\infty}^{+\infty} dp^- \Phi(p, P)_{\begin{subarray}{l} p^+=xP^+ \\  \end{subarray}}\,
= \int dp^-\, \text{Disc} [ \Phi(p, P) ]_{\begin{subarray}{l} p^+=xP^+ \\  \end{subarray}} 
\ , \ \ \ \ \ \ \ 
\Phi^{[\Gamma]}(x,p_T)\ \equiv\ \text{Tr} \bigg[ \Phi(x,p_T) \frac{\Gamma}{2} \bigg] \ ,
\eqno{(\theequation \textit{a,b})}
\end{equation*}
where the unintegrated $\Phi$ quark distribution correlator is defined as~\cite{Bacchetta:2006tn,Mulders:1995dh,Echevarria:2016scs}:
\begin{equation*}
\Phi_{ij}(p,P) = \int \frac{d^4 \xi}{(2\pi)^4} e^{\i p \cdot \xi}\, \text{Tr}_c 
\langle P|  
\timeord \big[ \overline{\psi}_j(0) W_2(0,\infty) \big] \, 
\antitimeord \big[ W_1(\infty,\xi) \psi_i(\xi) \big] \, 
|P \rangle  \, .
\end{equation*}
In the previous equations $p$ is the quark momentum, $P$ is the hadronic momentum, $x=p^+/P^+$ is the parton fractional momentum in the dominant direction. In Eq.~\eqref{e:proj_int_Phi}(b), the ``Tr'' operator ``Tr'' without subscripts indicates to a Dirac trace,  the $1/2$ factor is explained in Section 6.7 and 6.8 of Ref.~\cite{Collins:2011zzd}, and $\Gamma$ is a generic Dirac matrix;  for example, $\Gamma=\gamma^+$ is associated to the unpolarized TMD $f_1$, and the matrices associated to the other TMDs can be found in  \cite{Bacchetta:2006tn}.
In the spirit of Feynman rules, the Dirac trace operator ``$\text{Tr}$'' corresponds to the {\it sum} over the polarization states of the quark in the final state. 
The analogous color trace $\text{Tr}_c$, corresponding to a sum over the color configurations, is included in the definition of the correlator $\Phi$. 

Assuming that the correlators have the standard analiticity properties of the scattering amplitudes, the integration over the suppressed momentum component used to define the TMD correlators can be performed by complex contour deformation. Depending on the value of $x$, one can then replace the integral of the unintegrated correlator by the integral of its $s$-channel or $u$-channel discontinuity, denoted by ``Disc'' 
~\cite{Jaffe:1983hp,Boer:1998im,Diehl:2003ny,Gamberg:2010uw}. These correspond, respectively to a quark distribution ($0 \leq x \leq 1$), and to an antiquark distribution  ($-1 \leq x \leq 0$).

\subsection{Parton fragmentation correlator}

The TMD fragmentation correlator in the parton frame and its Dirac projections are defined as:

\begin{equation*}
\label{e:proj_int_Delta}
\refstepcounter{equation}
\Delta(z,P_T) \equiv  \int_{-\infty}^{+\infty} \frac{dk^+}{2z}\, \Delta(k, P)_{\begin{subarray}{l} P^-= zk^- \\  \end{subarray}}
= \int \frac{dk^+}{2z}\, \text{Disc}\, [ \Delta(k, P) ]_{\begin{subarray}{l} P^-= zk^- \\  \end{subarray}}
\ , \ \ \ \ \ \ \
\Delta^{[\Gamma]}(z,P_T)\ \equiv\ \text{Tr} \bigg[ \Delta(z,P_T) \frac{\Gamma}{2} \bigg] \ ,
\eqno{(\theequation \textit{a,b})}
\end{equation*}
with the unintegrated quark correlator $\Delta(k,P)$ defined in Eq.~\eqref{e:1hDelta_corr}. 
Here $k$ is the momentum of the fragmenting quark, $P$ is the momentum of the produced hadron, and $z=P^-/k^-$ is the hadron's fractional momentum in the dominant momentum direction. 
The $1/2$ factor in Eq.~\eqref{e:proj_int_Delta}(b) arises in the same way as in Eq.~\eqref{e:proj_int_Phi}(b). 
In Eq.~\eqref{e:proj_int_Delta}(a), the $1/z$ factor comes from the normalization of the hadronic states (see Ref.~\cite{Collins:1981uw} and Section 12.4 in Ref.~\cite{Collins:2011zzd}). 
The trace operator in this case has an additional $1/2$ factor which appears in Eq.~\eqref{e:proj_int_Delta}(a), since it corresponds to an {\it average} over the quark polarizations in the initial state. 
A color trace Tr$_c/N_c$, that in the same way corresponds to an average over the hadron's color configurations, is already included in the definition of the unintegrated correlator $\Delta(k,P)$. 

Note that in the fragmentation case we integrate over the suppressed partonic plus component even if the correlator has a probabilistic interpretation in terms of the hadronic variables. This is because the integration is always performed with respect to the momentum components of the object that in a process would enter the hard interaction in a process, namely the parton.

\subsection{Fully inclusive jet correlator}
In analogy with Eqs.~\eqref{e:proj_int_Delta}, the TMD inclusive jet correlator and its Dirac projections are defined as:
\begin{equation*}
\label{e:proj_int_J}
\refstepcounter{equation}
J(k^-,k_T)\, \equiv\ \frac{1}{2} \int dk^+  \, \Xi(k) 
\ , \ \ \ \ \ \ \ \ \
J^{[\Gamma]}(k^-,k_T)\, \equiv\  \text{Tr} \bigg[ J(k^-,k_T) \frac{\Gamma}{2} \bigg] \ .
\eqno{(\theequation \textit{a,b})}
\end{equation*}
For consistency with Ref.~\cite{Sterman:1986aj}, the discontinuity has been inserted directly in the definition of the unintegrated jet correlator~\eqref{e:invariant_quark_correlator}, or equivalently \eqref{e:invariant_quark_correlator_W}, where we also included the color trace Tr$_c/N_c$ corresponding to an average over the initial state color configurations. 
Note that there is no $1/k^-$ prefactor, at variance with Ref.~\cite{Accardi:2008ne}. 

\section{Frame transformations}
\label{a:frame_dep}


In this appendix we discuss the dependence of the fragmentation functions and of the associated momentum sum rules on the frame chosen to study the hadronization mechanism.

We will consider, in particular, two cases: the hadron frame used to define the TMD fragmentation functions \cite{Collins:2011zzd,Bacchetta:2006tn}, and the parton frame used in the main text to derive the momentum sum rules. Either frame is defined by a specific choice of basis four-vectors $n_{-(f)}$, $n_{+(f)}, n_{1(f)}, n_{2(f)}$, that identify the light-cone plus and minus directions and the two directions orthogonal to these, with an index $f=h,p$ explicitly referring to the hadron and parton frames, respectively. The basis vectors are not only utilized to define the corresponding coordinate system, but also to decompose the fragmentation correlator in terms of Fragmentation Functions, whose definition, consequently, depends on the choice of frame. In this appendix, therefore, we will consistently use the $h$ and $p$ subscripts to explicitly distinguish between quantities defined in one or the other frame. Note, however, that in the main text we dispensed from this notation.
\newline\indent
As discussed in Section~\ref{ss:1h_inclusive_FF}, the hadron frame is defined such that the momentum of the hadron under consideration has no transverse component ($\vect{P}_{Th}=0$); and the parton frame is such that the quark's momentum has no transverse component ($\vect{k}_{Tp}=0$). 
In the main body of this paper, we connected the fragmentation correlator and the quark propagator through a correlator-level sum rule that integrated over all hadron momenta. Hence only the quark's momentum is available to define the frame, and we could only choose to work in the parton frame. When discussing the fragmentation correlator, however, both frames are possible, with the hadron frame being the conventional choice~\cite{Collins:2011zzd,Bacchetta:2006tn}. As a consequence, the FFs functions defined in the TMD literature, and here generically denoted by $X_h$, differ from the fragmentation functions entering the sum rules for the parton-frame $X_p$ FFs summarized in Section~\ref{ss:1h_inclusive_FF}, where the $p$ index was dropped for simplicity. It is the purpose of this Appendix to derive the rules for transforming one set of FFs, and their corresponding sum rules and Equation of Motion relations, into the other. 

\subsection{Parton and hadron frames}
\label{ss:p_h_frames}

We will first consider the Lorentz transformation from the parton frame to the hadron frame. The transformation is completely determined by requiring that 
(1) the parton transverse momentum in the parton frame $\vect{k}_{Tp}$ be zero,  
(2) the minus component be invariant, and 
(3) the norm of any four-vector be invariant. 
The matrix associated to this transformation reads, in light-cone coordinates~\cite{Levelt:1993ac,Collins:2011zzd}:
\begin{equation}
  \label{e:p_to_h_frame}
  {\cal M}_{h \leftarrow p} = 
  \begin{bmatrix}
  1 & 0 & 0 \\
  \frac{\vect{k}_{Th}^2}{2(k^-)^2} & 1 & \frac{\vect{k}_{Th}}{k^-} \\
  \frac{\vect{k}_{Th}}{k^-} & 0 & 1
  \end{bmatrix}
  \,  ,
\end{equation}
where $\vect{k}_{Th}$ is the (Euclidean 2D) transverse momentum of the quark in the hadron frame. 
The hadron frame components $a_h^\mu$ of the vector $a$ can then be obtained from the parton frame components by
\begin{align}
   a_h^\mu = {({\cal M}_{h \leftarrow p})^\mu}_\nu \, a_p^\nu \ .
\end{align}
As the minus component is invariant, we will omit the subscript identifying the frame whenever little risk of misunderstanding occurs. The inverse transformation matrix ${\cal M}_{p \leftarrow h}$ from the hadron to the parton frame can be simply obtained by replacing $\vect{k}_{Th} \rightarrow - \vect{k}_{Th}$ in Eq.~\eqref{e:p_to_h_frame}.

From Eq.~\eqref{e:p_to_h_frame} one can see that the transverse momentum $\vect{P}_{Tp}$ of the hadron in the parton frame and the transverse momentum $\vect{k}_{Th}$ of the quark in the hadron frame are related by
\begin{equation}
\label{e:PTp_kTh_rel}
\vect{P}_{Tp} = -z\, \vect{k}_{Th} \, ,
\end{equation}
while the hadron's collinear momentum fraction relative to the quark is invariant between the two considered frames because of the invariance of the minus components of the momenta:
\begin{align}
\label{e:z_inv}
  z = \frac{P^-_p}{k^-_p} = \frac{P^-_h}{k^-_h} \ .
\end{align}

Let us now consider the transformation of the different ingredients in the Dirac decomposition of the TMD fragmentation correlator, that defines the fragmentation functions (see, for example, Eq.~\eqref{e:1hDelta_TMDcorr_param} for the decomposition in the parton frame).  
To start with, the metric tensor $g^{\mu\nu}$ is Lorentz-invariant by definition, and the Levi-Civita tensor $\epsilon^{\mu\nu\rho\sigma}$ is invariant, as well, since the transformation ${\cal M}_{h \leftarrow p}$ belongs to the orthochronus Lorentz group (det ${\cal M}_{h \leftarrow p} = 1$). 

The transformation of the transverse $g_T^{\mu\nu}$ and $\epsilon_T^{\mu\nu}$ tensors are, instead, more complex, and we need to first address the relation between the basis vectors defining the two frames under consideration.
Let's consider first the hadron frame basis vectors, which can be expressed in hadron-frame coordinates as ${n_-}_{(h)}^\mu = [1,0,\vect{0}_T]_h$, ${n_+}_{(h)}^\mu = [0,1,\vect{0}_T]_h$ and $n_{i(h)}^\mu = [0,1,\vect{e}_i]_h$, where $\vect{e}_1 = (1,0)$ and $\vect{e}_2=(0,1)$, and $i=1,2$ a transverse index. We also collect the transverse basis vectors $n_{i(h)}$ into a 2D transverse vector, $\vect{n}_{T(h)} \equiv (n_{1(h)},n_{2(h)})$. The parton frame basis vectors are analogously defined in parton frame coordinates. 
One can easily show that 
\begin{subequations}
\label{e:n_transf}
\begin{numcases}{}
n_{-(h)}^\mu = n_{-(p)}^\mu
  - \frac{1}{k^-} \vect{k}_{Th} \cdot \vect{n}_{T(p)}^\mu 
  + \frac12 \frac{\vect{k}_{Th}^2}{(k^-)^2} n_{+(p)}^\mu
  & \label{e:n_minus_transf} \\
n_{+(h)}^\mu = n_{+(p)}^\mu
  & \label{e:n_plus_transf} \\
\vect{n}_{T(h)}^\mu = \vect{n}_{T(p)}^\mu - \frac{\vect{k}_{Th}}{k^-} n_{+(p)}^\mu \ ,
  & \label{e:n_T_transf}     
\end{numcases}
\end{subequations}
It is then not difficult to obtain
\begin{align}
  \label{e:gT_transf}
g_{T(h)}^{\mu\nu} & = g_{T(p)}^{\mu\nu} + \frac{1}{k^-} \vect{k}_{Th} \cdot \vect{n}_{T(p)}^{\{\mu} n_{+(p)}^{\nu\}} - \frac{\vect{k}_{Th}^2}{(k^-)^2} n_{+(p)}^\mu n_{+(p)}^\nu\\
  \label{e:epsT_trans}
\epsilon_{T(h)}^{\mu\nu} & = \epsilon_{T(p)}^{\mu\nu} - \frac{1}{k^-} \epsilon^{\mu\nu\rho\sigma} \vect{k}_{Th} \cdot \vect{n}_{T(p)\rho} \, n_{+(p)\sigma} \ .
\end{align}
While neither tensor is actually Lorentz invariant in itself, the breaking terms are at least of $O(1/k^-)$.
In our application to the Lorentz  transformation of the fragmentation correlator $\Delta$, we only need these tensors contracted with $k_\nu$, with a much simpler transformation: 
\begin{align}
\label{e:nTdot_transf}
g_{T(h)}^{\mu\nu} k_\nu & = -\frac{1}{z} g_{T(p)}^{\mu\nu} P_\nu - \frac{\vect{P}_{Tp}^2}{z^2k^-} n_{+(p)}^\mu  
\\
\label{e:epsTdot_transf}
\epsilon_{T(h)}^{\mu\nu} k_\nu & = -\frac{1}{z} \epsilon_{T(p)}^{\mu\nu} P_\nu \ .
\end{align}
Note that Eq.~\eqref{e:nTdot_transf} generalizes Eq.~\eqref{e:PTp_kTh_rel} to the four vector case. 
Finally, the first four Dirac matrices transform as any other four-vector: 
\begin{subequations}
\label{e:gammamu_transf}
\begin{numcases}{}
\gamma_h^- = \gamma_p^- \equiv \gamma^- 
& \label{e:gammaminus_transf} \\
\gamma_h^+ = \gamma_p^+ 
+ \frac{\vect{k}_{Th}^2}{2(k^-)^2}\, \gamma^- 
+ \frac{\vect{k}_{Th} \cdot \vect{\gamma}_{Tp}}{k^-}  
& \label{e:gammaplus_transf} \\
\vect{\gamma}_{Th} = \vect{\gamma}_{Tp} + \frac{\vect{k}_{Th}}{k^-}\, \gamma^- \ ,
& \label{e:gammaT_transf}     
\end{numcases}
\end{subequations}
with $\gamma_5 = \varepsilon_{\mu\nu\rho\sigma} \gamma^\mu \gamma^\nu\gamma^\rho \gamma^\sigma$ invariant because such is the Levi-Civita tensor and we are working in 4 dimensions \cite{Collins:1984xc}.

We now have all the tools to understand what happens to the fragmentation functions, the Equation of Motion relations (EOMs), and the momentum sum rules when changing frames.

\subsection{Transformation of the fragmentation functions}
\label{ss:FFtransforms}

As discussed, the TMD fragmentation functions are conventionally defined by decomposing the TMD fragmentation correlator in terms of the light cone basis vectors of the hadron frame~\cite{Bacchetta:2006tn}: 
\begin{align}
\label{e:1hDelta_TMDcorr_param_hadron_frame}
{\Delta}_{(h)}(z,k_{T(h)}) & = 
\frac{1}{2} \slashed{n}_{-(h)} D_{1h}(z,k_{Th}^2) + 
\i \frac{ \big[ \slashed{k}_{T(h)}, \slashed{n}_{-(h)} \big]}{4 M} H_{1h}^{\perp}(z,k_{Th}^2) + 
\frac{M}{2 P^-} E_h(z,k_{Th}^2) \\
\nonumber & 
+ \frac{\slashed{k}_{T(h)}}{2 P^-} D_h^{\perp}(z,k_{Th}^2) 
+ \frac{\i M}{4 P^-} \big[ \slashed{n}_{-(h)}, \slashed{n}_{+(h)} \big] H_h(z,k_{Th}^2) 
+ \frac{1}{2 P^-} \gamma_5\, \epsilon_{T(h)}^{\rho\sigma}\, \gamma_{\rho}\, {k_\sigma} \, G_h^{\perp}(z,k_{Th}^2) \ .
\end{align}
Note that we omitted the frame subscript for the frame-independent quantities, and that $\slashed{k}_{T(h)} = \gamma_\mu\, g_{T(h)}^{\mu\nu} k_\nu$. We have also dropped the flavor index on the mass $M$ to avoid confusion with the frame subscript $h$, and used $k_{Th}^2=k_{T(h)}^\mu k\big._{T(h)\mu}$ as a shorthand in the argument of the fragmentation functions. 

It is now important to realize that the TMD correlator $\Delta_{(h)}(z,k_{Th}^2) \equiv (2z)^{-1} \int dk_h^+ \Delta(k,P)$ is invariant under Lorentz trasformations, such as the hadron to parton frame transformation under discussion in this appendix, that connect frames with the same light-cone plus axis. Explicitly,
\begin{align}
\label{e:Delta_h_vs_p}
    \Delta_{(p)}(z,P_{Tp}^2) =  \Delta_{(h)}(z,k_{Th}^2) \ ,
\end{align}
where the TMD correlator $\Delta_{(p)} = (2z)^{-1} \int dk_p^+ \Delta(k,P)$ is decomposed in terms of the parton-frame light-cone basis vectors, see Eq.~\eqref{e:1hDelta_TMDcorr_param} in the main text. This can be seen in two steps. First, notice that the integration over $dk^+$ is Lorentz invariant because the minus and transverse components are fixed. Then, look at the definition \eqref{e:1hDelta_ampl} of the unintegrated $\Delta^h(k,P)$: on the one hand, there are no open Lorentz indexes; on the other hand, the light-cone plus vectors associated  with the $B_i$ functions is the same in the considered frames.

%

Schematically, the Dirac projections that define the FFs take the form
\begin{equation}
\label{e:inv_traces}
  X_f \sim \text{Tr}\Big[ \Delta_{\text{TMD}}\, \Gamma_{(f)} \Big] \ ,
\end{equation}
where $\Gamma_{(f)}$ is a suitable contraction of the Dirac matrices and the light-cone basis vectors for a given frame $f$, see for example 
Eqs.~\eqref{e:sumrule_D1},~\eqref{e:sumrule_E},~\eqref{e:sumrule_H},~\eqref{e:sumrule_H1p},~\eqref{e:sumrule_Dp},~\eqref{e:sumrule_Gp}.  
When  performing a Lorentz transformation, one needs to keep all the involved vectors unchanged. Under this condition, the traces in Eq.~\eqref{e:inv_traces} are Lorentz invariant. If one changes the basis vectors, though, one obtains a {\em different} definition of fragmentation functions, and one can study how these different fragmentation functions transform into one another.

Specifically, let's consider the transformation between the hadron-frame FFs introduced in this appendix, and the parton-frame FFs discussed in the main text. This can be obtained by decomposing, with the help of Eq.~\eqref{e:n_transf}, the $n_{(h)}$ vectors in Eq.~\eqref{e:Delta_h_vs_p} on the parton-frame light-cone basis, and transforming the transverse momentum components according to Eq.~\eqref{e:PTp_kTh_rel}. The parton frame FFs can then be projected out utilizing the $\Gamma_{(p)}$ functions, or, more simply, obtained by matching the corresponding Dirac structures in Eq.~\eqref{e:Delta_h_vs_p}. One finds 
\begin{align}
\label{e:Dperp_transf}
& D_p^\perp(z,P_{Tp}^2) = D_h^\perp(z,k_{Th}^2) - z\, D_{1h}(z,k_{Th}^2) \equiv {\widetilde D}_h^\perp(z,k_{Th}^2) \, , \\
\label{e:H_transf} 
& H_p(z,P_{Tp}^2) = H_h(z,k_{Th}^2) -z\, \frac{k_{Th}^2}{M^2}\, H_{1h}^\perp(z,k_{Th}^2) \equiv {\widetilde H}_h(z,k_{Th}^2) \,  ,
\end{align}
while all other fragmentation functions do not mix\footnote{For the $G^\perp$function, this is actually true only when summing over the hadron spins.}:
\begin{align}
  \label{e:p_to_h_genericFF}
  X_p (z,P_{Tp}^2) = X_h(z,k_{Th}^2) \ .
\end{align}
In practice, the change of basis vectors mixes the twist-2 FFs in the hadron frame ($D_{1h}$ and $H_{1h}^\perp$) with two other twist-3 FFs ($D_h^\perp$ and $H_h$) through the off-diagonal terms in Eq.~\eqref{e:p_to_h_frame} proportional to $1/k^- \sim 1/P_h^-$. The other FFs do not mix under this change of basis.  

The identification of the r.h.s. of Eqs.~\eqref{e:Dperp_transf} and~\eqref{e:H_transf} with the ${\widetilde D}_h$ and ${\widetilde H}_h$ functions requires one to use the hadron frame EOM relations discussed in Ref.~\cite{Bacchetta:2006tn}. It is important to remark that these tilde-functions are among the functions that parametrize the dynamical twist-3 quark-gluon-quark correlator $\Delta_A^\alpha$~\cite{Bacchetta:2006tn}. Hence, Eqs.~\eqref{e:Dperp_transf} and~\eqref{e:H_transf} imply that the distinction between kinematical and dynamical twist-3 is, for certain functions, frame-dependent, and the transformation ${\cal M}_{h \leftarrow p}$ actually maps a kinematical twist-3 quantity into a dynamical one. A similar version of the transformation~\eqref{e:Dperp_transf} for $D^\perp$ was already discussed in Ref.~\cite{Levelt:1993ac}, whereas the transformation~\eqref{e:H_transf} for $H$ is, to our knowledge, new.

Before moving to collinear functions, it is worthwhile remarking an important, but potentially confusing difference between our notation (derived e.g. from Ref.~\cite{Bacchetta:2006tn}) and that of, {\em e.g.}, Refs~\cite{Metz:2016swz,Collins:2011zzd}. In the hadron frame, the natural transverse momentum variable for a FF is $k_{Th}$, as we have used in this Appendix. However, the physical interpretation of a FF should be given in the partonic frame. Hence, reading Eq.~\eqref{e:p_to_h_genericFF} from right to left, and utilizing Eq.~\eqref{e:PTp_kTh_rel}, we find 
\begin{align}
     X_h(z,k_{Th}^2) = X_p (z,z^2 k_{Th}^2)\ .
\end{align}
In other words, the hadron frame FFs depend on the $z^2 k_{Th}^2$ combination, rather than $k_{Th}^2$ alone. This justifies using $z^2 k_{Th}^2$ as argument of $X_h$ as done in e.g.  Refs~\cite{Metz:2016swz,Collins:2011zzd}. An important consequence is that, if one wishes to use a Gaussian approximation for the transverse momentum dependence of the FFs in the hadron frame, this should read $X_h(z,k_{Th}^2) \approx D(z) \exp\big[-z^2 \vect{k}_{Th}^2/(\Delta^2)\big]$, where $\Delta^2$ is the variance, and $D$ a function of $z$ alone. 

The collinear FFs are usually defined in the parton frame as integrals of the TMD FFs over the transverse momentum (see Refs.~\cite{Bacchetta:2006tn,Metz:2016swz}). The definition in the hadron frame follows, if one requires the collinear FFs to be frame independent. 
Explicitly,
\begin{equation}
\label{e:D1_coll_def}
X_{p}(z) \equiv \int d^2 \vect{P}_{Tp}\, X_{p}(z,P_{Tp}^2)   \, ,
\quad \quad \quad 
X_{h}(z) \equiv z^2 \int d^2 \vect{k}_{Th}\, X_{h}(z,k_{Th}^2) \ ,
\end{equation}
and it is easy to see that 
\begin{align}
  \label{e:coll_FF_lorentz}   
  X_{p}(z)=X_{h}(z)
 \end{align}
for FFs that do not mix under Lorentz transformations.
Following the standard conventions discussed in Refs.~\cite{Bacchetta:2006tn,Metz:2016swz}, the first transverse moments are defined in the parton and hadron frames as
\begin{equation}
\label{e:first_mom} 
X_p^{(1)}(z) \equiv \int d^2 \vect{P}_{Tp}\, \frac{\vect{P}_{Tp}^2}{2 z^2 M^2}\, X_p(z,P_{Tp}^2) \, ,
\quad \quad \quad 
X_h^{(1)}(z) \equiv z^2\, \int d^2 \vect{k}_{Th}\, \frac{\vect{k}_{Th}^2}{2M^2}\, X_h(z,k_{Th}^2) \, .
\end{equation} 
These also do not mix, $X^{(1)}_{p}(z)=X^{(1)}_{h}(z)$, except for the $D^\perp$ and $H$ functions.

\subsection{Transformation of the EOMs}

The EOMs allow one to relate twist-2 fragmentation functions with kinematical and dynamical twist-3 fragmentation functions. 
They can be obtained by applying the Dirac equation $(i\slashed{D}(\xi) - m)\psi(\xi)=0$ to the fragmentation correlator and projecting on the good quark components. The resulting relation between the twist-2 and twist-3 fragmentation correlators is Lorentz covariant, as seen in Eq. (3.53) of Ref.~\cite{Bacchetta:2006tn}, which is given in the hadron frame.  
Using the transformation~\eqref{e:p_to_h_frame} it is possible to show that the EOMs are frame invariant up to terms suppressed by powers of $M/P^-$, which can be kinematically neglected in a frame boosted to high values of $P^-$ such as we are considereing in this paper~\cite{Bacchetta:2006tn}. 
Thus, in order to derive the EOMs in the parton frame, one simply needs to apply the replacement rule~\eqref{e:PTp_kTh_rel} for the transverse momenta to the hadron frame EOMs given in Ref.~\cite{Bacchetta:2006tn}:  
\begin{equation}
  \label{e:eom_p_frame}
  \begin{aligned}
  & E_p = \widetilde{E}_p + z \frac{m_{q0}}{M} D_{1p} 
  &\quad\quad& D_p^{\perp} = \widetilde{D}_p^{\perp} + z D_{1p} \\
  & H_p = \widetilde{H}_p - \frac{\vect{P}_{Tp}^2}{z M^2}   H_{1p}^\perp
  && G_p^{\perp} = \widetilde{G}_p^{\perp} + z \frac{m_{q0}}{M}   H_{1p}^{\perp\,} \ .
  \end{aligned}
\end {equation}
These are the EOMs utilized in Section~\ref{ss:qgq_sumrules}, but are written with an explicit $p$ frame subscript.

\subsection{Transformation of the sum rules}

We now discuss the frame (in)dependence of the momentum sum rules for the fragmentation functions. 

Let us start from the rank $0$ sum rules for the $D_1$ and $E$ fragmentation functions, derived in Section~\ref{ss:mu_minus} in the parton frame.  
The Dirac matrices that project these functions out of the fragmentation correlator are, respectively, the $\gamma^-$ and $\id$ matrices. Since these are invariant under the ${\cal M}_{h \leftarrow p}$ transformation, also the momentum sum rules for $D_1$ and $E$ are invariant. 

The transformation of the rank $0$ parton frame sum rule for $H$ is less straightforward, because of its mixing with the Collins function $H_1^\perp$, and it is instructive to look at the latter first. The projection matrix for $H_1^\perp$ is $\Gamma = \i \sigma^{i-}\gamma_5$, which renders the RHS of the master sum rule~\eqref{e:sumrule_transverse} equal zero. Moreover this matrix is frame independent,  because the extra term from Eq.~\eqref{e:gammaT_transf} cancels in the commutator that defines $\sigma^{i-}$. Accordingly, the rank $1$ sum rule for $H_1^\perp$ is Lorentz-invariant. As a consequence of Eq.~\eqref{e:H_transf}, also the sum rule for $H$ is, despite the fact that this FF mixes with the Collins function under Lorentz transformations. 

The other two FFs appearing in the rank $1$ sum rules discussed in Section~\ref{ss:mu_transverse} are $G^\perp$ and $D^\perp$. 
The projection matrix for $G^\perp$ is $\Gamma = \gamma_T^i\, \gamma_5$, such that the RHS of the sum rule is zero, as well. Differently from the sum rule for the Collins function, this matrix does get an extra term proportional to $\gamma^- \gamma_5$ under Lorentz transformation, which is however only related to polarized FFs. Since the sum rules can only be obtained after summing over the hadronic polarizations, this extra term does not contribute and the sum rule for $G^\perp$ is Lorentz invariant. 

The sum rule for $D^\perp$ in the parton frame involves the hadronic transverse momentum rescaled by the collinear momentum fraction averaged over the kinematics and summed over the produced hadrons and their spin. 
It is therefore useful to introduce the notion of the average of a momentum-dependent $O_p=O_p(z,P_{Tp}^2)$ observable in the parton frame  as:
\begin{equation}
\label{e:def_average_pframe}
\langle O_p \rangle = \sum_{H, S} \int dz\, d^2 \vect{P}_{Tp}\, O_p(z,P_{Tp}^2)\, z\, D_{1}^{H}(z,P_{Tp}^2) \ ,
\end{equation}
where, at variance with the main text, we used an upper case hadronic $H$ flavor index to distinguish this from the hadronic frame index (this definition can also be extended to a flavor dependent observable $O^H$, but we suppressed that index for clarity).  
The average operator is Lorentz invariant if we define this for a hadron frame observable $O_h=O_h(z,k_{Th}^2)$ as
\begin{equation}
\label{e:def_average_hframe}
\langle O_h \rangle \equiv \sum_{H, S} \int dz\, d^2 \vect{k}_{Th}\, O_h(z,k_{Th}^2)\, z^3\, D_{1}^{H}(z,k_{Th}^2) \, .
\end{equation}
Now, one can calculate the hadron frame sum rule for $D^\perp$ by applying the ${\cal M}_{h \leftarrow p}$ Lorentz transformation and the mixing relation~\eqref{e:Dperp_transf} to the parton frame sum rule~\eqref{e:pframe_sumrule_Dp_Dpt}$a$. 
Utilizing the EOMs in the two frames, it is also possible to obtain the hadron frame sum rule for $\widetilde{D}^\perp$. The result of these manipulation is given in Table~\ref{t:sum_rules_Dperp},  where we collect and compare the $D^\perp$ and $\widetilde{D}^\perp$ sum rules in either frame, expressed in terms of the average  defined in Eqs.~\eqref{e:def_average_pframe} and~\eqref{e:def_average_hframe}.

\begin{table}
\small
 \centering
\begin{tabular}{|c||c|c|}
  \hline
  \multicolumn{1}{|c||}{\ frame\ \ } & 
  \multicolumn{1}{c|}{\ $\displaystyle 2 \sum_{H,S} \int dz\, M_H^2\, {D}_f^{H\perp(1)}(z)$\ \ } & 
  \multicolumn{1}{c|}{\ $\displaystyle 2 \sum_{H,S} \int dz\, M_H^2\, \widetilde{D}_f^{H\perp(1)}(z)$\ \ } \\
  \hline
  \hline
          &   &  \\ 
$f = p$ & 0 & $\langle P_{Tp}^2/z^2 \rangle$ \\              
          &   &  \\ 
$f =  h$ & $\langle \vect{k}_{Th}^2 \rangle$ & 0  \\                      
          &   &  \\ 
\hline
\end{tabular}
\caption{Sum rules for the $D^\perp$ and $\widetilde{D}^\perp$ twist-3 FFs in the parton frame ($f = p$) and in the hadron frame ($f =  h$). The flavor index is denoted by an uppercase $H$ to distinguish this from the lowercase $h$ frame index.}
\label{t:sum_rules_Dperp}
\end{table} 

One can notice an interesting symmetry between the results obtained in the parton and in the hadron frames. 
In the parton frame, the twist-3 $\widetilde{D}^\perp$ sum rule measures the average squared hadronic transverse momentum dynamically generated during the hadronization process scaled by a factor $1/z$, while the $D^\perp$ sum rule is trivial. 
In the hadron frame, instead, it is the twist-2 sum rule for $D^\perp$ that measures the dynamical generation of transverse momentum. In this case the averaged quantity is formally the transverse partonic momentum as seen by the hadron. 
While different in form, the two averages measure the same quantity, as is obvious from Eq.~\eqref{e:PTp_kTh_rel}. The formal frame dependence of the $D^\perp$ and $\widetilde D^\perp$ sum rules is a consequence of the  fact that these FFs enter the fragmentation correlator with a coefficient proportional to the transverse momentum, and of the Lorentz transformation properties of the latter.

\end{appendix}

\bibliographystyle{JHEP}  
\bibliography{JSR.bib}

\end{document}